\documentclass[11pt]{article}
\usepackage{cite}
\usepackage{scalerel}
\usepackage{xcolor}
\usepackage{shuffle}
\usepackage{amsmath,amsfonts,amssymb}
\usepackage{latexsym,epsfig}
\usepackage{dsfont}
\usepackage{hyperref}
\usepackage{extarrows}
\usepackage{comment} 
\usepackage{tikz}
\usetikzlibrary{decorations.pathmorphing}
\usepackage{mathtools}
\tikzset{snake it/.style={decorate, decoration=snake}}

\newcommand{\mycomment}[1]{}

\def\hybrid{
        \topmargin -20pt
        \oddsidemargin 0pt
        \headheight 0pt \headsep 0pt
        \textwidth 6.25in 
        \textheight 9.5in 
        \marginparwidth .875in
        \parskip 5pt plus 1pt \jot = 1.5ex}

\hybrid

 \csname
@addtoreset\endcsname{equation}{section}

\newcommand{\be}{\begin{equation}}
\newcommand{\ee}{\end{equation}}

\newcommand{\ba}{\begin{equation} \begin{aligned}}
\newcommand{\ea}{\end{aligned} \end{equation}}

\def\cD{\mathcal{D}}

\def\cA{{\cal A}}

\def\cN{{\cal N}}

\def\cU{{\cal U}}
\def\cV{{\cal V}}

\def\del{\partial}
\def\l{\langle}
\def\r{\rangle}

\def\B{\square}

\allowdisplaybreaks

\def\bpm{\begin{pmatrix}}
\def\epm{\end{pmatrix}}
\newcommand{\bra}[1]{\langle#1\rvert}
\newcommand{\ket}[1]{\lvert#1\rangle}

\begin{document}

\begin{titlepage}
\rightline{}
\rightline{February  2025}
\rightline{HU-EP-25/11-RTG}  
\begin{center}
\vskip 1.5cm
{\Large \bf{Worldline geometries for scattering amplitudes}}

\vskip 1.7cm

{\large\bf {Roberto Bonezzi, Maria Foteini Kallimani}}
\vskip 1.6cm

{\it  Institute for Physics, Humboldt University Berlin,\\
 Zum Gro\ss en Windkanal 2, D-12489 Berlin, Germany}\\[1.5ex] 
 roberto.bonezzi@physik.hu-berlin.de,\\ maria.foteini.kallimani@physik.hu-berlin.de
\vskip .1cm

\vskip .2cm

\end{center}

\bigskip\bigskip
\begin{center} 
\textbf{Abstract}

\end{center} 
\begin{quote}

In this paper, we construct the path integral for infinite and semi-infinite scalar worldlines. We show that, at the asymptotic endpoints, on-shell physical states can be generated  by inserting vertex operators at infinity. This procedure implements automatically the LSZ reduction, thus leading to a direct worldline representation of scattering amplitudes. To obtain it, we introduce generalized vertex operators, to be viewed as the gluing of entire tree subdiagrams to a given worldline. We demonstrate that the subdiagrams themselves are given, via a recursive relation, by correlation functions on the semi-infinite line. In this sense, the approach we take is fully first-quantized, in that it does not need any field theoretic quantity as input. We envisage that, when suitably extended to gauge theories, it could provide useful insights in addressing current research issues, such as color-kinematics duality.

\end{quote} 
\vfill
\setcounter{footnote}{0}
\end{titlepage}

\tableofcontents

\section{Introduction}

In the modern history of theoretical physics there are two, seemingly unrelated, approaches to compute relativistic scattering amplitudes. The quantum field theory (QFT) textbook approach is based on the perturbative expansion of Green's functions and amplitudes in terms of Feynman diagrams.
On the other hand, scattering amplitudes can also be computed using first-quantized methods, mostly known from string theory. In the latter case, one computes target space amplitudes from correlation functions of vertex operators on worldsheets of different topologies. These two approaches are to some extent complementary: the string path integral only computes amplitudes with on-shell physical states, but it does so by repackaging a multitude of Feynman diagrams in a single worldsheet correlator. For this reason, the worldsheet representation of amplitudes often makes some properties more transparent: most prominently, string perturbation theory displays a better organization of gauge invariance, and it was instrumental to reveal the hidden relations between gauge theory and gravity amplitudes \cite{Kawai:1985xq} that sparked the entire double copy program \cite{Bern:2008qj,Bern:2019prr,Bern:2022wqg,Adamo:2022dcm}.

The worldline approach to quantum field theory sits somewhat in between, in that it employs first-quantized techniques akin to string theory, to describe processes involving conventional point particles. The idea of treating particles in a scattering process by relativistic worldline path integrals dates back to the seminal work of Feynman \cite{Feynman:1950ir,Feynman:1951gn} which, however, was soon superseded by the advent of modern quantum field theory. This point of view was revived in the early nineties, when Strassler \cite{Strassler:1992zr}, building on the work of Bern and Kosower on the point particle limit of string amplitudes \cite{Bern:1990cu,Bern:1992cz}, developed the worldline techniques that are still used nowadays. In this framework, a propagating scalar or spinor particle is treated as a relativistic worldline extending between two spacetime points. The emission or absorption of quanta, such as photons \cite{DHoker:1995uyv,Schubert:2001he,Ahmadiniaz:2020wlm,Ahmadiniaz:2021gsd}, gluons \cite{Bastianelli:2013pta,Ahmadiniaz:2015xoa} or gravitons \cite{Bastianelli:2002fv,Bastianelli:2002qw}, is described by inserting suitable vertex operators along the line, in a way that closely resembles string perturbation theory. Later on, the worldline formalism has been extended to include gluons in a non-abelian background \cite{Dai:2008bh,Bonezzi:2024emt} and gravitons on curved spacetime \cite{Bonezzi:2018box,Bastianelli:2019xhi,Bastianelli:2022pqq,Bastianelli:2023oyz,Bastianelli:2023oca}, with the aim of providing a first-quantized description of self-interacting Yang-Mills theory and gravity. Most recently, it has found a prominent application to the problem of classical gravitational scattering \cite{Mogull:2020sak,Jakobsen:2021zvh,Haddad:2024ebn}, while also serving as a promising framework to investigate color-kinematics duality \cite{Ahmadiniaz:2021fey,Ahmadiniaz:2021ayd,Bonezzi:2024fhd}, thanks to its similarity to first-quantized string theory. 

Compared to string theory, the worldline formalism has the advantage of making direct contact with QFT calculations, not having to deal with the infinite tower of massive string states. Likewise, since the quantized worldline is not constrained by conformal invariance, it can easily describe off-shell quantities, such as correlation functions, without the severe complications of string field theory. This being said, there are some evident shortcomings, which we start addressing in this paper focusing on tree-level processes. Contrary to string theory, a worldline correlator does account for several distinct Feynman diagrams, but not for all diagrams needed for a complete amplitude. More specifically, inserting a number of vertex operators into an open worldline generates only so-called half-ladder diagrams, with external legs sprouting from a main propagation line. This misses all diagrams where entire subtrees are attached to the main line. While these do not appear in QED, where photon emissions from scalar or spinor lines give the full dressed propagator, they are necessary for gauge invariance in QCD and pure Yang-Mills theory. Usually, these contributions are added by hand employing some appropriate ``pinching rules'' \cite{Bern:1990cu,Bern:1992cz}. These result in a hybrid representation, making use of both QFT and worldline objects, which is not ideal to exploit the string-like properties of the first-quantized formulation. A second, related, drawback is that the worldline endpoints and the states represented by vertex operators are treated on a very different footing, while in a self-interacting field theory, as well as in string theory, all external lines are equivalent. In particular, the standard path integral of the open worldline describes the (possibly dressed) propagator between two spacetime points. To obtain the contribution to a scattering amplitude, one has to go through the process of Fourier transforming the positions of the endpoints, and implementing the LSZ reduction via the usual QFT methods \cite{Daikouji:1995dz,Mogull:2020sak,Ahmadiniaz:2021gsd}.

In this paper, we propose a resolution to these issues for the tree level of self-interacting scalar theories. Ultimately, the goal will be to extend the ideas presented here to the more interesting case of non Abelian gauge theories, but the scalar case already displays the relevant geometric concepts and provides the skeletal structure of the worldline perturbation theory. For the open worldline, we identify three distinct geometries: the finite, infinite and semi-infinite lines. We show that these correspond to different objects in the field theory, those being dressed propagators, on-shell amplitudes and multiparticle fields, respectively. In the following, we will refer to the standard relativistic path integral with position eigenstates at the endpoints as the finite line. This is the one mostly studied in the literature, although there are some examples where different boundary conditions are considered, see for instance \cite{Dai:2008bh,Mogull:2020sak,Bonocore:2020xuj}. 

Here we give a systematic derivation of the path integral for the infinite and semi-infinite lines. For both cases, the proper length of the line is infinite and they differ in that the semi-infinite one extends from a fixed spacetime position $x^\mu$ to infinity. The three cases are then distinguished by the number of boundaries: two for the finite line, one for the semi-infinite and none for the infinite. We give boundary conditions to the path integral, so that at infinite times it projects onto the zero-momentum eigenstate, while asymptotic states are created by placing vertex operators at infinity. This prescription not only simplifies the evaluation of correlation functions, but it also makes closest contact with string theory, since all external states are treated on equal footing as vertex operators. Additionally, we show that the states created at asymptotically large times are forced to be on shell, thus implementing the LSZ reduction automatically.

The central building block of our construction is the semi-infinite worldline. As a core result of this work, we prove that its path integral with insertions of vertex operators generates nonlinear multiparticle fields of arbitrary order via the recursion
\begin{equation}\label{intro recursion}
\sum_{m=1}^\infty \frac{1}{(m-1)!}\phi^{(m)}(x)=\prescript{}{x}{\!\!\left\langle\exp\left(\sum_{k=1}^\infty \frac{1}{k!}\overline{V}^{(k)}\right)\phi^{(1)}(X(\infty))\right\rangle}\;,
\end{equation}
where $\overline{V}^{(k)}$ are generalized vertex operators, to be defined in the main text, and angular brackets denote the normalized path integral average on the semi-infinite line, with fixed boundary at $x^\mu$.
These multiparticle fields are represented by all rooted tree diagrams, in which external legs are on-shell states and the root is the fixed spacetime point $x^\mu$ where the field is evaluated. The recursive formula \eqref{intro recursion} can be viewed as a first-quantized version of the Berends-Giele recursion for Yang-Mills currents \cite{Berends:1987me}, or of the so-called perturbiner expansion \cite{Mafra:2016ltu,Mizera:2018jbh,Lopez-Arcos:2019hvg}. We further introduce the concept of generalized vertex operators, which are viewed as the gluing of subtrees, given by multiparticle fields, to a worldline. Putting to fruition these novel perspectives, we provide the worldline representation of arbitrary tree-level amplitudes, which are obtained as correlation functions of generalized vertex operators on the infinite line. 
We view the results of this paper as a clear indication that the worldline formalism is more flexible than it is usually believed. In particular, in our construction, every object is obtained as a worldline path integral and does not invoke ad hoc inputs from quantum field theory at any stage. In this regard, it can be considered as a genuine first-quantized approach which, if suitably extended to non Abelian gauge theories, could provide valuable insights into their elusive kinematic algebra, building on the results of \cite{Ahmadiniaz:2021fey,Ahmadiniaz:2021ayd,Bonezzi:2024fhd}.

The rest of the paper is organized as follows: in Section \ref{sec:review} we give a quick review of the path integral quantization of the scalar relativistic particle, focusing on the standard derivation of vertex operators and dressed propagators. Section \ref{sec:wline geometries} is devoted to construct the path integral and generating functionals for the less explored cases of the infinite and semi-infinite worldlines. There, we also start to introduce the idea of multiparticle fields and generalized vertex operators as correlation functions on the semi-infinite line, restricting to the case of $\phi^3$ theory. These concepts are made precise in Section \ref{sec:recursions}, where we prove the recursive formula \eqref{intro recursion} to all orders and extend it to the case of arbitrary polynomial interactions. We close in Section \ref{sec:conclusions} with our conclusions. 

\section{Relativistic particle and scalar field theory}\label{sec:review}

In this section we will briefly review the path integral quantization of the scalar relativistic particle, emphasizing its relation to various quantities in scalar field theories. We start from the free particle on an open worldline and subsequently introduce a scalar potential. Its path integral is related to the field theoretic dressed propagator, from which one can define the standard vertex operators for the emission-absorption of scalar quanta.

\subsection{Scalar relativistic particle}

The standard action for a free relativistic particle in Minkowski spacetime is just the proper time of the worldline trajectory. In most cases, however, it is more convenient to work with the ``Polyakov'' form of the action, which makes use of an intrinsic einbein $e(\tau)$. The first order, or Hamiltonian, action is given in terms of the phase space variables $(X^\mu,P_\mu)$ and the einbein $e$
\begin{equation}\label{SfreeHam}
S[X,P,e]=\int_0^1 d\tau\,\Big[P_\mu\dot X^\mu-e\left(P^2+m^2\right)\Big]\;,    
\end{equation}
where $\mu=0,\ldots, D-1$ is a spacetime Lorentz index.
The action is invariant under time reparametrizations, with local parameter $\xi(\tau)$:
\begin{equation}
\delta X^\mu=2\xi P^\mu\;,\quad\delta P^\mu=0\;,\quad\delta e=\dot\xi   \;. 
\end{equation}
On $X^\mu$ and $P_\mu$ the above transformations are generated canonically by the Hamiltonian constraint $H=P^2+m^2$ and are equivalent to one-dimensional diffeomorphisms. 
The first order form of the action makes manifest the role of the einbein as the Lagrange multiplier for the mass-shell constraint, which is part of the equations of motion
\begin{equation}
P^\mu=\frac{1}{2e}\,\dot X^\mu\;,\quad \dot P^\mu=0\;,\quad P^2+m^2=0\;.    
\end{equation}
One can integrate out the momentum $P^\mu$ using its algebraic equation, yielding the second order action
\begin{equation}\label{SfreeLag}
S[X,e]=\int_0^1 d\tau\,\left[\,\frac{1}{4e}\,\dot X^2-em^2\right]\;.   
\end{equation}
This is still invariant under the local transformations
\begin{equation}\label{LagrangianDiffeo}
\delta X^\mu= \xi\dot X^\mu\;,\quad\delta e=\del_\tau(e\xi)\;,   
\end{equation}
which now take the form of standard diffeomorphisms for a one-dimensional scalar and one-form, respectively.

We now turn to the path integral quantization of the relativistic particle. To do so, we have to specify the topology of the worldline and to deal with the above gauge symmetry. In this paper we will discuss only open worldlines. We thus specify boundary conditions for the trajectory $X^\mu(\tau)$ at the endpoints by fixing
\begin{equation}
X^\mu(0)=x^\mu\;,\quad X^\mu(1)=y^\mu\;.    
\end{equation}
In order to preserve the above boundary conditions, the diffeomorphism parameter has to vanish at the endpoints: $\xi(0)=\xi(1)=0$. This also ensures that the action is gauge invariant without boundary terms. Given that $\xi(\tau)$ vanishes at the endpoints, the integral
\begin{equation}
T:=\int_0^1d\tau\,e(\tau)\;,    
\end{equation}
is a gauge invariant modulus. One can thus use the gauge symmetry \eqref{LagrangianDiffeo} to fix $e(\tau)=T$, but the path integral over the einbein has a leftover integration over the modulus $T$.
Following the Faddeev-Popov procedure, the path integral is schematically given by
\begin{equation}
\int\frac{\cD X\cD e}{\rm VolGauge}\,e^{iS[X,e]}=\int_0^\infty dT\int \cD X\,e^{iS[X,T]}\int \cD c\cD b\,e^{iS_{\rm gh}} \;,   
\end{equation}
where the ghost action is $S_{\rm gh}=\int d\tau\,b\dot c$. The ghost $c(\tau)$ inherits the vanishing boundary conditions of the diffeomorphism parameter $\xi(\tau)$. This implies that the antighost admits a zero-mode $b_0$, which drops from the action. As in string theory, where antighost zero-modes are related to metric moduli, the zero-mode of $b$ corresponds to the modulus $T$. A more careful treatment of the modulus indeed shows that the ghost path integral does contain a $b_0$ insertion, which allows us to normalize it to one: $\int \cD c\cD b\,b_0\,e^{iS_{\rm gh}}=1$. 

Factoring out the ghost path integral, we are left with the following partition function for the open line:
\begin{equation}
\label{free dd heat kernel}
\begin{split}
Z_0(x,y)&=\int_0^\infty dT\,e^{-im^2T}K_0(x,y;T)\;,\\
K_0(x,y;T)&=\int\big[\cD X\big]_x^y\exp\left\{i\int_0^Td\tau\,\frac{\dot X^2}{4}\right\}\;,
\end{split}
\end{equation}
where we have extracted the constant mass term and rescaled the worldline parameter as $\tau\to\frac{1}{T}\tau$. The functional measure $\big[\cD X\big]_x^y$ denotes integration over paths with boundary conditions $X^\mu(0)=x^\mu$, $X^\mu(T)=y^\mu$. The free heat kernel $K_0(x,y;T)$ is formally identical to a non-relativistic path integral with Hamiltonian $P^2$ and propagation time $T$. To compute it, we first split the trajectory $X^\mu(\tau)$ into a classical path obeying the boundary conditions, plus quantum fluctuations:
\begin{equation}\label{XsplitDD}
X^\mu(\tau)=x^\mu+(y^\mu-x^\mu)\,\frac{\tau}{T}+z^\mu(\tau)\;,\quad z^\mu(0)=z^\mu(T)=0\;.    
\end{equation}
The integration is now performed over the fluctuation $z^\mu(\tau)$ with measure $\big[\cD X\big]_x^y=\big[\cD z\big]_{\rm VBC}$, where VBC stands for vanishing boundary conditions.
The heat kernel factorizes into the exponential of the on-shell action times a $T$-dependent constant, given by the free path integral over $z^\mu(\tau)$, as 
\begin{equation}
K_0(x,y;T)=e^{\frac{i(x-y)^2}{4T}}\cN(T)\;,\quad\cN(T)=\int\big[\cD z\big]_{\rm VBC} \exp\left\{i\int_0^Td\tau\,\frac{\dot z^2}{4}\right\} \;.  
\end{equation}
The normalization $\cN(T)$ can be fixed by matching the functional integral with the appropriate transition amplitude in operator language. The fluctuation $z^\mu$ with vanishing boundary conditions corresponds to initial and final position eigenstates with vanishing eigenvalue, denoted by $\ket{x=0}$, which allows us to set
\begin{equation}
\cN(T)=\bra{x=0}e^{-iT\hat P^2}\ket{x=0}=\frac{1}{(4\pi iT)^{D/2}}\;,    
\end{equation}
upon inserting a completeness relation in terms of momentum eigenstates $\mathds{1}=\int\frac{d^Dk}{(2\pi)^D}\ket{k}\bra{k}$, normalized as $\l x|k\r=e^{ik\cdot x}$.

The relativistic partition function is then given by 
\begin{equation}
Z_0(x,y)=\int_0^\infty\frac{dT}{(4\pi iT)^{D/2}}\,e^{i\frac{(x-y)^2}{4T}-im^2T}    \;,
\end{equation}
which coincides with the field theory propagator for a massive scalar, written in position space.
To obtain the familiar momentum space expression, we rewrite the heat kernel as a Gaussian Fourier transform
\begin{equation}
K_0(x,y;T)=\frac{e^{\frac{i(x-y)^2}{4T}}}{(4\pi iT)^{D/2}}=\int\frac{d^Dp}{(2\pi)^D}\,e^{-ip^2T+ip\cdot(x-y)} \;,   
\end{equation}
and perform the modular integral, yielding
\begin{equation}
\label{dirichlet dirichlet path integral}
Z_0(x,y)=\int\frac{d^Dp}{(2\pi)^D}\,e^{ip\cdot(x-y)}\int_0^\infty dT\,e^{-i(p^2+m^2)T}=-i\int\frac{d^Dp}{(2\pi)^D}\,\frac{e^{ip\cdot(x-y)}}{p^2+m^2-i\epsilon}\;,    
\end{equation}
where the Feynman $i\epsilon$ prescription arises as a regulator for the integral over $T$. Notice that, since $T$ has dimensions of inverse mass square, the integration regions $T\simeq0$ and $T\to\infty$ correspond to the ultraviolet and infrared regions of the field theory, respectively.

\subsection{Particle in a background field and dressed propagator}

So far we have reviewed how the path integral of the free particle on the line describes the scalar propagator in free field theory. One can add an arbitrary function of $X^\mu$ to the Hamiltonian constraint in \eqref{SfreeHam}, while retaining reparametrization invariance. The second order action
\begin{equation}\label{SVLag}
S[X,e]=\int_0^1 d\tau\,\Big[\,\frac{1}{4e}\,\dot X^2-e\,\cV(X)\Big]\;,   
\end{equation}
is still invariant under worldline diffeomorphisms \eqref{LagrangianDiffeo} for any function $\cV(X)$, which can contain the constant mass term. The gauge fixing procedure is carried out as in the free theory, leading to the partition function 
\begin{equation}\label{Zxygenericpotential}
Z_\cV(x,y)=\int_0^\infty dT\int\big[\cD X\big]_x^y\exp\left\{i\int_0^Td\tau\left(\frac14\,\dot X^2-\cV(X)\right)\right\} \;,   
\end{equation}
for the open line with fixed endpoints. The heat kernel now corresponds to the non-relativistic Hamiltonian $H=P^2+\cV(X)$, with propagation time $T$.
For generic potentials $\cV(X)$ the above path integral can only be evaluated perturbatively.

To appreciate the meaning of the partition function \eqref{Zxygenericpotential}, let us consider a scalar field theory with spacetime action
\begin{equation}
S[\Phi]=\int d^Dx\Big[-\frac12\,\del^\mu\Phi\del_\mu\Phi-\cU(\Phi)\Big] \;,   
\end{equation}
where $\cU(\Phi)$ is an arbitrary polynomial potential. In the background field method, one expands $\Phi=\phi+\eta$, where $\phi$ is the classical background field and $\eta$ the quantum fluctuation. The quadratic action for $\eta$ then reads
\begin{equation}
S_2[\eta;\phi]=\frac12\,\int d^Dx\,\eta\big(\B-\cU''(\phi)\big)\eta \;.   
\end{equation}
The formal inverse of the effective kinetic operator is the so-called dressed propagator for $\eta$, which contains all possible  insertions of the background field $\phi$. In position space it is defined by the matrix element
\begin{equation}
D(x,y;\phi)=\bra{y}\frac{i}{\B-\cU''(\phi)}\ket{x} \;,   
\end{equation}
which can be exponentiated using Schwinger's parametrization $\frac{1}{A}=\int_0^\infty ds\,e^{-sA}$. The resulting matrix element can be interpreted as a quantum mechanical transition amplitude, namely
\begin{equation}
D(x,y;\phi)=\int_0^\infty dT\,\bra{y}e^{-i\big(\hat P^2+\,\cU''(\phi)\big)T}\ket{x} \;,
\end{equation}
where we identified $-i \del_\mu\equiv\hat P_\mu$. The above transition amplitude can finally be represented as a non-relativistic path integral with classical Hamiltonian $H=P^2+\cU''\big(\phi(X)\big)$. Including the integration over Schwinger's proper time, this shows that the partition function \eqref{Zxygenericpotential} computes the dressed propagator
\begin{equation}
D(x,y;\phi)\equiv Z_\cV(x,y)\;,    
\end{equation}
where the worldline potential $\cV$ is related to the field theory one by $\cV(X)=\cU''\big(\phi(X)\big)$.

\subsection{Vertex operators and Feynman diagrams}\label{sec:2.3}

We continue the discussion of the worldline representation of the dressed propagator by specializing to the case of massive $\phi^3$ theory. The field theory potential is given by
\begin{equation}
\cU(\Phi)=\frac12\,m^2\Phi^2+\frac{1}{3!}\,g\,\Phi^3\quad\longrightarrow\quad\cU''(\phi)=m^2+g\,\phi\;,    
\end{equation}
corresponding to the quantum mechanical potential $\cV(X)=m^2+g\,\phi(X)$. Isolating again the constant mass term, the dressed propagator coincides with the partition function
\begin{equation}\label{Zxyphi3}
\begin{split}
Z_\phi(x,y)&=\int_0^\infty dT\,e^{-im^2T}K_\phi(x,y;T) \;,\\
K_\phi(x,y;T)&=\int\big[\cD X\big]_x^y\exp\left\{i\int_0^Td\tau\left(\frac14\,\dot X^2-g\,\phi(X)\right)\right\}\;.
\end{split}   
\end{equation}

The heat kernel $K_\phi(x,y;T)$ can only be computed in some perturbative scheme. For instance, one can compute it systematically as a power series in $T$. This results in an expansion in terms of so-called Seeley-DeWitt coefficients, which control the short $T$ behavior of the heat kernel and, in turn, the ultraviolet properties of the dressed propagator. In field theory, typically, the relevant perturbative expansion is in powers of the coupling $g$. The path integral \eqref{Zxyphi3}, however, cannot be computed order by order in $g$ for arbitrary background fields $\phi(x)$. To do so, one usually considers field configurations that describe a fixed number of free particles. This leads, as we will review next, to the concept of vertex operators.

In order to discuss how vertex operators emerge, we start manipulating the path integral into a perturbative expansion in $g$. As in the free theory, we split the trajectory into a classical path plus fluctuations: $X^\mu(\tau)=x^\mu_{\rm cl}(\tau)+z^\mu(\tau)$, where
\begin{equation}\label{x class plus z}
x_{\rm cl}^\mu(\tau)=x^\mu+(y^\mu-x^\mu)\,\frac{\tau}{T}\;,\quad z^\mu(0)=z^\mu(T)=0\;.    
\end{equation}
The classical path satisfies the boundary conditions, but it is only a solution of the free equations of motion. Inserting the split \eqref{x class plus z} into the action yields
\begin{equation}
\int_0^Td\tau\left[\frac{\dot X^2}{4}-g\,\phi(X)\right]=\frac{(x-y)^2}{4T}+\int_0^Td\tau\left[\frac{\dot z^2}{4}-g\,\phi(x_{\rm cl}+z)\right]\;.    
\end{equation}
Upon defining the normalized average  with respect to the free $z$ path integral
\begin{equation}\label{zaverageDD}
\big\l F[z]\big\r:=(4\pi iT)^{D/2}\int\big[\cD z\big]_{\rm VBC}\,F[z]\,\exp\left\{i\int_0^Td\tau\frac{\dot z^2}{4}\right\}\;,\quad\l1\r=1 \;,   
\end{equation}
the heat kernel \eqref{Zxyphi3} can be written as a power series in $g$ of worldline correlators
\begin{equation}\label{Heatphi power series}
\begin{split}
K_\phi(x,y;T)&=\frac{e^{\frac{i(x-y)^2}{4T}}}{(4\pi iT)^{D/2}}\,\left\l e^{-ig\int_0^Td\tau\phi(\tau)}\right\r\\
&=\frac{e^{\frac{i(x-y)^2}{4T}}}{(4\pi iT)^{D/2}}\,\sum_{n=0}^\infty\frac{(-ig)^n}{n!}\int_0^Td\tau_1\cdots\int_0^Td\tau_n\,\big\l\phi(\tau_1)\cdots\phi(\tau_n)\big\r\;,
\end{split}    
\end{equation}
where we used the shorthand notation $\phi(\tau):=\phi\big(x_{\rm cl}(\tau)+z(\tau)\big)$.

It is clear that no order in $g$ can be computed explicitly for a generic function $\phi(x)$. Rather, at fixed order $g^n$, one usually computes the contribution to the dressed propagator corresponding to the interaction with $n$ particles of momenta $k_1,\ldots,k_n$. We denote such contribution as $D_n(x,y;\{k_i\})$. In terms of standard Feynman diagrams, this is given by all half-ladder diagrams with $n$ external legs (either on-shell or off-shell) attached to the main propagator line. At order $g^n$, the field configuration giving rise to these diagrams is a sum of plane waves
\begin{equation}
\phi(x)=\sum_{i=1}^ne^{ik_i\cdot x} \;,   
\end{equation}
where in \eqref{Heatphi power series} one keeps only the multilinear contribution, with each plane wave appearing once. This leads to defining the vertex operator
\begin{equation}
V_i(\tau):=-ig\,e^{ik_i\cdot X(\tau)}\;, \quad X^\mu(\tau)=x^\mu_{\rm cl}(\tau)+z^\mu(\tau)\;,   
\end{equation}
which describes the emission (or absorption) of the $i$-th scalar particle with momentum $k_i^\mu$ by the worldline, at time $\tau$.

With this notation at hand, the worldline representation for $D_n(x,y;\{k_i\})$ is given in terms of the correlation function of $n$ integrated vertex operators, which is very similar to the way one computes scattering amplitudes in string theory. Upon using \eqref{Heatphi power series} in the partition function \eqref{Zxyphi3}, one finally obtains
\begin{equation}\label{Dressed propagator}
D_n(x,y;\{k_i\})=\int_0^\infty dT\,e^{-im^2T}\frac{e^{\frac{i(x-y)^2}{4T}}}{(4\pi iT)^{D/2}}\,\int_0^Td\tau_1\cdots\int_0^Td\tau_n\,\Big\l V_{1}(\tau_1)\cdots V_{n}(\tau_n)\Big\r\;.    
\end{equation}
The main difference with the formula \eqref{Heatphi power series} for a generic background field is that the correlator of an arbitrary number of vertex operators can be computed exactly. We will come back to this in the next section. 
We illustrate diagrammatically the path integral \eqref{Dressed propagator} as:
\begin{center}
    \begin{tikzpicture}[scale=2]
        \node at (-1,0) {$D_n(x,y;\{k_i\})=$};
        \filldraw (0,0) circle (1pt);
        \draw (0,0)--(3,0);
        \filldraw (3,0) circle (1pt);
        \draw[snake it] (1,0)--(1,1);
        \draw[snake it] (2,0)--(2,1);
        \node at (0,-0.3) {\small{$X^\mu(0)=x^\mu$}};
        \node at (3,-0.3) {\small{$X^\mu(T)=y^\mu$}};
        \node at (1,-0.3) {\small{$\tau_1$}};
        \node at (2,-0.3) {\small{$\tau_n$}};
        \node at (1.5, 0.3) {$\dots$};
        \node at (0.8,1) {$k_1$};
        \node at (2.2,1) {$k_n$};
    \end{tikzpicture}
\end{center}
where the fixed worldline endpoints are depicted by black dots. The wavy lines denote insertions of integrated vertex operators $V_{i}(\tau_i)$ with definite momenta.

We conclude the review of the standard worldline approach to scalar trees, exemplified for the case of $\phi^3$ theory, with a couple of remarks:
\begin{itemize}
\item In order to use $D_n(x,y;\{k_i\})$ to compute the corresponding contribution to an $(n+2)$-point scattering amplitude, one has to first Fourier transform it in the endpoint positions $x$ and $y$ and then put the resulting legs on shell by performing the LSZ reduction.
\item The dressed propagator $D_n(x,y;\{k_i\})$ only encodes half-ladder diagrams, but it misses all diagrams where nontrivial subtrees are attached to the main worldline. 
\end{itemize}
Given this state of affairs, in the next sections we will give a direct path integral representation for scattering amplitudes, bypassing the steps of the LSZ reduction. We will then introduce generalized vertex operators, whose aim is to treat all tree-level diagrams on equal footing.

\section{Open worldline geometries}\label{sec:wline geometries}

If one is ultimately interested in computing scattering amplitudes, there are three relevant geometries for the open worldline: finite, infinite and semi-infinite. Their path integrals have different field-theoretic interpretations and, correspondingly, are associated with different boundary conditions for the worldline fields. We will discuss them in turn, focusing on the appropriate boundary conditions and related generating functionals. 

\subsection{Finite line: dressed propagators}

The first case we are interested in is the standard path integral with fixed boundary conditions for the trajectory $X^\mu(\tau)$, which we have reviewed in the previous section. As we have mentioned, the Dirichlet boundary conditions force the diffeomorphism parameter $\xi(\tau)$ to vanish at both endpoints. This, in turn, implies that the proper length of the line $T$ is a gauge invariant modulus. Since the partition function is given by the $T$-integral of the transition amplitude $K(x,y;T)$ for fixed propagation time, we refer to this geometry as the finite line, or Dirichlet-Dirichlet line, reflecting the boundary conditions
\begin{equation}
X^\mu(0)=x^\mu\;,\quad X^\mu(T)=y^\mu\;.    
\end{equation}
In the field theory context, this path integral produces a (possibly dressed) scalar propagator, where the initial and final states are necessarily off shell.

As we have seen in the previous section, the perturbative evaluation of the path integral reduces to computing correlation functions of the fluctuation $z^\mu(\tau)$ defined in \eqref{XsplitDD}. To this end, we introduce the generating functional
\begin{equation}
\begin{split}
Z_{\rm DD}[j]&:=\int\big[\cD z\big]_{\rm VBC}\,\exp\left\{i\int_0^Td\tau\left(\frac14\,\dot z^2+j_\mu z^\mu\right)\right\}\\
&=\frac{1}{(4\pi iT)^{D/2}}\,\left\l e^{i\int_0^Td\tau\,j(\tau)\cdot z(\tau)}\right\r_{\rm DD}\;,
\end{split}
\end{equation}
in terms of the normalized average \eqref{zaverageDD}, where we have added the subscript DD as a reminder of the boundary conditions. The free correlation functions are given in terms of the worldline propagator $G_{\rm DD}(\tau,\sigma)$ by
\begin{equation}
\begin{split}
\left\l\exp\,i\int_{0}^{T}d\tau\,j_\mu(\tau)\,z^\mu(\tau)\right\r_{\rm DD}&=\exp\left\{\frac{i}{2}\int_{0}^{T}d\tau d\sigma\,j^\mu(\tau)\,G_{\rm DD}(\tau,\sigma)\,j_\mu(\sigma)\right\}\;,\\
G_{\rm DD}(\tau,\sigma)&=|\tau-\sigma|-(\tau+\sigma)+\frac{2}{T}\,\tau\sigma\;,
\end{split}    
\end{equation}
so that $\l z^\mu(\tau)z^\nu(\sigma)\r_{\rm DD}=-i\,\eta^{\mu\nu}G_{\rm DD}(\tau,\sigma)$. The above propagator is the Green's function for $\frac12\del_\tau^2$ vanishing at both $\tau,\sigma=0$ and $\tau,\sigma=T$. In particular, the boundary conditions break translation invariance and fix $G_{\rm DD}(\tau,\sigma)$ completely.

This result for the generating functional can be used to compute the correlation function of an arbitrary number of vertex operators. For instance, the correlator appearing in \eqref{Dressed propagator} yields
\begin{equation}
\begin{split}
\Big\l V_{1}(\tau_1)\cdots V_{n}(\tau_n)\Big\r_{\rm DD}&=(-ig)^ne^{i\sum_{i=1}^nk_i\cdot x_{\rm cl}(\tau_i)}\Big\l e^{i\sum_{j=1}^nk_j\cdot z(\tau_j)}\Big\r_{\rm DD}\\
&=(-ig)^ne^{i\sum_{i=1}^nk_i\cdot x_{\rm cl}(\tau_i)}e^{\frac{i}{2}\sum_{i,j}k_i\cdot k_jG_{\rm DD}(\tau_i,\tau_j)}\;,
\end{split}
\end{equation}
and we recall that the classical trajectory is defined as $x_{\rm cl}^\mu(\tau)=x^\mu+(y^\mu-x^\mu)\,\frac{\tau}{T}$.
As prescribed by \eqref{Dressed propagator}, all vertex operators have to be integrated over the worldline.
This type of master formulas is the usual starting point for computing dressed propagators in the worldline formalism. In view of obtaining on-shell scattering amplitudes, we argue that this Dirichlet-Dirichlet path integral is needed only beyond one loop, where the worldline endpoints $x$ and $y$ can be glued to other lines to form higher loops. At tree-level, on the other hand, we find it more transparent to consider a different worldline geometry, which allows us to obtain directly on-shell amplitudes in momentum space.

\subsection{Infinite line: tree-level amplitudes}\label{sec:3.2}

In the following we will construct the path integral for a worldline of infinite length, which carries momentum eigenstates at asymptotic times. We will show that the path integral is non-vanishing only if the asymptotic states are on shell, thus making direct contact with scattering amplitudes. 

\subsubsection*{Path integral for momentum eigenstates}

Since the path integral with momentum eigenstates at the boundaries is less explored 
in the worldline literature, we give a thorough discussion of its generating functional, before studying the case of the infinite line.
For a simpler comparison with the operator formulation, we find it convenient to start from the first-order action
\begin{equation}\label{HamAction}
S[X,P]=\int_0^Td\tau\,\Big[-\dot P_\mu X^\mu-H(X,P)\Big]\;,     
\end{equation}
with boundary conditions
\begin{equation}\label{Pboundary conditions}
P_\mu(0)=k_\mu\;,\quad P_\mu(T)=k'_\mu\;.    
\end{equation}
The symplectic structure ensures that the variational principle is well-defined for trajectories with fixed boundary conditions in $P_\mu(\tau)$ and differs from the standard form $P_\mu\dot X^\mu$ by boundary terms.
The action \eqref{HamAction} can be viewed as the relativistic action with Hamiltonian constraint $H(X,P)$, upon gauge fixing the einbein $e=T$ and rescaling $\tau\rightarrow\tau/T$. The relevant Hamiltonian for applications to $\phi^3$ theory is then $H(X,P)=P^2+m^2+g\,\phi(X)$.

We define the phase space path integral with boundary conditions \eqref{Pboundary conditions} by matching it with the transition amplitude between momentum eigenstates
\begin{equation}
\bra{k'}e^{-iT\hat H}\ket{k}=\int \cD X\big[\cD P\big]_k^{k'}\,e^{iS[X,P]}\;.   
\end{equation}
In order to fix the normalization of the path integral, let us specialize
to the massless free theory with $H=P^2$. The amplitude is proportional to a delta function, since momentum is conserved:
\begin{equation}\label{operator transition}
\bra{k'}e^{-iT\hat P^2}\ket{k}=(2\pi)^D\delta^D(k-k')\,e^{-iTk^2}\;.
\end{equation}
To extract a normalizable path integral, we first separate the zero-mode from the coordinates:
\begin{equation}
X^\mu(\tau)=x_0^\mu+z^\mu(\tau)\;,\quad\int_0^Td\tau\,z^\mu(\tau)=0\;.    
\end{equation}
The quadratic action then splits into a boundary term plus the bulk action for $z^\mu$ and $P_\mu$
\begin{equation}
\begin{split}
S_2[X,P]&=\int_0^Td\tau\,\Big[-\dot P_\mu X^\mu-P^2\Big]=(k-k')\cdot x_0+\int_0^Td\tau\,\Big[-\dot P_\mu z^\mu-P^2\Big]\\
&=(k-k')\cdot x_0+S_2[z,P]\;.
\end{split}    
\end{equation}
The integration measure for $X$ factorizes as $\cD X=d^Dx_0\cD z$ and the integral over the zero-mode produces the delta function
\begin{equation}
\begin{split}
\int \cD X\big[\cD P\big]_k^{k'}\,e^{iS_2[X,P]}&=\int d^Dx_0\,e^{i(k-k')\cdot x_0}\int \cD z\big[\cD P\big]_k^{k'}\,e^{iS_2[z,P]}\\
&=(2\pi)^D\delta^D(k-k')\,\int \cD z\big[\cD P\big]_k^{k'}\,e^{iS_2[z,P]}\;.
\end{split}    
\end{equation}
Matching this with the operator result \eqref{operator transition} and integrating over $k'$ fixes the normalization for the free path integral, provided that the initial and final momenta coincide and $z^\mu$ has no zero-mode
\begin{equation}
\int \cD z\big[\cD P\big]_k^{k}\,e^{iS_2[z,P]}=e^{-iTk^2}\;.    
\end{equation}
In particular, from now on we will take the momentum to have vanishing boundary conditions: $k_\mu=k_\mu'=0$, in which case the above path integral is normalized to one.

To compute expectation values of functionals of $z^\mu$ between eigenstates with \emph{zero momentum}, we take $P_\mu(0)=P_\mu(T)=0$ and couple $z^\mu$ to a source $j'_\mu(\tau)$ without a zero-mode, i.e. obeying $\int_0^Td\tau\,j'_\mu(\tau)=0$. The resulting normalized generating functional is given by
\begin{equation}\label{Z genpp}
\begin{split}
\left\l e^{i\int_0^Td\tau\,j'(\tau)\cdot z(\tau)}\right\r_{\rm NN}&:=\int \cD z\big[\cD P\big]_{\rm VBC}\,\exp\left\{i\int_0^Td\tau\left(-\dot P_\mu z^\mu-P^2+j'_\mu z^\mu\right)\right\}\\
&=\exp\left\{\frac{i}{2}\int_0^Td\tau d\sigma\,j'^{\mu}(\tau)\,G_{\rm NN}(\tau,\sigma)\,j'_\mu(\sigma)\right\}\;,
\end{split}    
\end{equation}
where the NN subscript refers to vanishing Neumann boundary conditions at both endpoints, although they are really vanishing Dirichlet conditions for momentum.
The propagator
\begin{equation}\label{G Neumann}
G_{\rm NN}(\tau,\sigma)=|\tau-\sigma|-\frac{1}{T}\,(\tau^2+\sigma^2)+\tau+\sigma-\frac{2T}{3}    \;,
\end{equation}
obeys $\frac12\del_\tau^2G_{\rm NN}(\tau,\sigma)=\delta(\tau-\sigma)-1/T$, where the right-hand side is the appropriate identity operator in the space of functions with no zero-mode. Accordingly, $G_{\rm NN}(\tau,\sigma)$ has vanishing integral in both $\tau$ and $\sigma$. On the other hand, the derivative $\del_\tau G_{\rm NN}(\tau,\sigma)$ vanishes for both $\tau=0$ and $\tau=T$. This reflects correctly the boundary conditions of $P_\mu(\tau)$, since its equation of motion $P^\mu=\frac12\,\dot z^\mu$ relates it to the derivative of $z^\mu$.

We can exploit this further and integrate out the momentum in \eqref{Z genpp}. This can be done without generating boundary terms, owing to the vanishing boundary conditions of $P_\mu$. The generating functional \eqref{Z genpp} can thus be expressed via the second-order path integral
\begin{equation}\label{Normalized DzPI}
\begin{split}
\left\l e^{i\int_0^Td\tau\,j'(\tau)\cdot z(\tau)}\right\r_{\rm NN}&=\int \big[\cD z\big]'_{\rm NN}\,\exp\left\{i\int_0^Td\tau\left(\frac14\,\dot z^2+j'_\mu z^\mu\right)\right\}\\
&=\exp\left\{\frac{i}{2}\int_0^Td\tau d\sigma\,j'^{\mu}(\tau)\,G_{\rm NN}(\tau,\sigma)\,j'_\mu(\sigma)\right\}\;,
\end{split}    
\end{equation}
where the measure $\big[\cD z\big]'_{\rm NN}$ denotes vanishing Neumann boundary conditions at both endpoints and the absence of the zero-mode, i.e.
\begin{equation}
\dot z^\mu(0)=\dot z^\mu(T)=0\;,\quad \int_0^Td\tau\,z^\mu(\tau)=0\;.    
\end{equation}
We can relax the non-local constraint on $z^\mu$ by reintroducing the zero-mode $x_0^\mu$ and working with the trajectory $X^\mu(\tau)=x^\mu_0+z^\mu(\tau)$ subject only to the vanishing Neumann conditions. We further couple the zero-mode $x_0^\mu$ to a source $j_{\mu0}$, so that we can define an unconstrained current $j_\mu(\tau)=j'_\mu(\tau)+\frac{1}{T}\,j_{\mu0}$. We are thus led to define the Neumann-Neumann path integral
\begin{equation}\label{Z Neumann}
\begin{split}
Z_{\rm NN}[j]&:=\int \big[\cD X\big]_{\rm NN}\,\exp\left\{i\int_0^Td\tau\left(\frac14\,\dot X^2+j_\mu X^\mu\right)\right\}\\
&=\int d^Dx_0\,e^{ix_0\cdot j_0}\int \big[\cD z\big]'_{\rm NN}\,\exp\left\{i\int_0^Td\tau\left(\frac14\,\dot z^2+j'_\mu z^\mu\right)\right\}\\
&=(2\pi)^D\delta^D(j_0)\,\left\l e^{i\int_0^Td\tau\,j'(\tau)\cdot z(\tau)}\right\r_{\rm NN}\;,
\end{split}    
\end{equation}
where the normalized expectation value is given by \eqref{Normalized DzPI}.

With the construction presented so far one can compute correlation functions between zero-momentum eigenstates, but we are eventually interested in eigenstates with initial momentum $k$ and final momentum $k'$. Rather than representing these states by changing boundary conditions, we will create them by inserting vertex operators at the initial and final times, thanks to the operator relation
\begin{equation}\label{momentum shift}
\ket{k}=e^{ik\cdot \hat X}\ket{0}\;,    
\end{equation}
where $\hat X^\mu$ is the quantum mechanical position operator and $\ket{0}$ the eigenstate with zero momentum. To test this idea, let us consider the expectation value of three vertex operators
\begin{equation}
\bra{0}e^{-ik'\cdot \hat X(T)}e^{ip\cdot \hat X(\tau_*)}e^{ik\cdot \hat X(0)}\ket{0}\;,
\end{equation}
written with the position operators in the Heisenberg picture: $\hat X^\mu(\tau)=e^{i\hat H\tau}\hat X^\mu e^{-i\hat H\tau}$. We take $\tau_*$ to be an intermediate time $0<\tau_*<T$, so that the correlator is time ordered. In a free theory with $\hat H=\hat P^2$ we can easily compute it by using \eqref{momentum shift} and $\hat P^2\ket{0}=0$, yielding
\begin{equation}\label{operator test}
\begin{split}
\bra{0}e^{-ik'\cdot \hat X(T)}e^{ip\cdot \hat X(\tau_*)}e^{ik\cdot \hat X(0)}\ket{0}&=\bra{0}e^{-ik'\cdot \hat X}e^{-i\hat P^2(T-\tau_*)}e^{ip\cdot \hat X}e^{-i\hat P^2\tau_*}e^{ik\cdot \hat X}\ket{0}\\
&=\bra{k'}e^{-i\hat P^2(T-\tau_*)}e^{ip\cdot \hat X}e^{-i\hat P^2\tau_*}\ket{k}\\
&=(2\pi)^D\delta^D(k+p-k')\,\exp\left\{-i\left(\tau_*k^2+(T-\tau_*){k'}^{2}\right)\right\}\;.
\end{split}    
\end{equation}

According to our previous discussion, the above result should be reproduced by the path integral with three insertions, namely
\begin{equation}\label{path test}
\bra{0}e^{-ik'\cdot \hat X(T)}e^{ip\cdot \hat X(\tau_*)}e^{ik\cdot \hat X(0)}\ket{0}=\int \big[\cD X\big]_{\rm NN}\,e^{-ik'\cdot X(T)}e^{ip\cdot X(\tau_*)}e^{ik\cdot X(0)}\,e^{i\int_0^Td\tau\,\frac14\,\dot X^2}\;.    
\end{equation}
The functional integral can be computed exactly, since it coincides with the generating functional \eqref{Z Neumann} with source $j_\mu(\tau)=k_\mu\,\delta(\tau)+p_\mu\,\delta(\tau-\tau_*)-k'_\mu\,\delta(\tau-T)$, resulting in
\begin{equation}\label{path test almost}
\begin{split}
\int \big[\cD X\big]_{\rm NN}\,e^{-ik'\cdot X(T)}e^{ip\cdot X(\tau_*)}e^{ik\cdot X(0)}\,e^{i\int_0^Td\tau\,\frac14\,\dot X^2}&=(2\pi)^D\delta^D(k+p-k')\\
&\times\left\l e^{-ik'\cdot z(T)}e^{ip\cdot z(\tau_*)}e^{ik\cdot z(0)}\right\r_{\rm NN} \;.   
\end{split}
\end{equation}
Evaluating the normalized correlator with \eqref{Normalized DzPI} and \eqref{G Neumann} we find
\begin{equation}\label{path test success}
\begin{split}
\left\l e^{-ik'\cdot z(T)}e^{ip\cdot z(\tau_*)}e^{ik\cdot z(0)}\right\r_{\rm NN}
&=\exp\Big\{\frac{i}{2}\,\Big(k^2G_{\rm NN}(0,0)+p^2G_{\rm NN}(\tau_*,\tau_*)+{k'}^2G_{\rm NN}(T,T)\\
&-2\,p\cdot k'G_{\rm NN}(\tau_*,T)-2\,k\cdot k'G_{\rm NN}(0,T)+2\,k\cdot p\,G_{\rm NN}(0,\tau_*)\Big)\Big\}\\
&=\exp\Big\{-i\,\Big(\tau_*k^2+(T-\tau_*){k'}^2\Big)\Big\}\;,
\end{split}    
\end{equation}
upon using momentum conservation, which reproduces the operator result \eqref{operator test}.

With a finite line, the relativistic path integral requires to further integrate the vertex operator at $\tau_*$ over the entire worldline, as well as to integrate over the proper time $T$. Integrating over all possible propagation times, the initial and final states are off shell and we still obtain a contribution to the dressed propagator, this time fully in momentum space. To see this, we evaluate explicitly the integral of \eqref{path test success} for the massive case, which requires to add the phase factor $e^{-im^2T}$, cf. \eqref{Dressed propagator}. Including also a factor of $-ig$ for the vertex operator at $\tau_*$ we obtain
\begin{equation}
\begin{split}
&-ig\int_0^\infty dT\,e^{-im^2T}\int_0^Td\tau_*\left\l e^{-ik'\cdot z(T)}e^{ip\cdot z(\tau_*)}e^{ik\cdot z(0)}\right\r_{\rm NN}\\
&=-ig\int_0^\infty\!\!\! dT\int_0^T\!\!\!d\tau_*\,e^{-i(k^2+m^2)\tau_*}e^{-i({k'}^2+m^2)(T-\tau_*)}=-ig\int_0^\infty\!\!\! ds\int_0^\infty\!\!\! d\tau_*\,e^{-i(k^2+m^2)\tau_*}e^{-i({k'}^2+m^2)s}\\
&=\frac{ig}{\big(k^2+m^2\big)\big((k+p)^2+m^2\big)}\;,
\end{split}    
\end{equation}
where we recall that $k'=k+p$ due to the momentum-conserving delta function in \eqref{path test almost}.
This is exactly the contribution with one scalar insertion to the dressed propagator in $\phi^3$ theory. In particular, notice that the scalar leg associated with the vertex operator at $\tau_*$ is amputated, but not necessarily on shell.

\subsubsection*{The infinite line and scattering amplitudes}

Having derived the path integral for momentum eigenstates, we can now begin to explore the geometry of the infinite line. To this end, it is best to revisit the gauge fixing procedure of the relativistic action. We start from the reparametrization invariant action
\begin{equation}
S[X,e]=\int_{-1/2}^{1/2}d\tau\,\left[\,\frac{1}{4e}\,\dot X^2-em^2\right]\;,    
\end{equation}
where we translated the range of $\tau$ to be symmetric around zero. Notice that in this form the affine parameter $\tau$ is dimensionless, while the einbein $e$ has dimensions of inverse mass square. 
Since the intrinsic length of the worldline is
$T:=\int_{-1/2}^{1/2}d\tau\,e(\tau)$, declaring that the line is infinite amounts to the limit $T\rightarrow\infty$. With a dynamical einbein, however, any statement about the finiteness of the line is equivalent to a global condition on the allowed $e(\tau)$ in the path integral. 

To better understand this point,  
we first rescale $\tau\rightarrow\tau/T$, $e\rightarrow Te$ and then take the limit $T\rightarrow\infty$. The action becomes 
\begin{equation}\label{Infinite action}
S[X,e]=\int_{-\infty}^{+\infty}d\tau\,\left[\,\frac{1}{4e}\,\dot X^2-em^2\right]\;,    
\end{equation}
where now the einbein is dimensionless and $\tau$ has dimensions of inverse mass square. Under a one-dimensional diffeomorphism, with $\delta X^\mu=\xi\dot X^\mu$ and $\delta e=\del_\tau(e\xi)$, the Lagrangian changes by a total derivative, so that
\begin{equation}
\delta S[X,e]=\int_{-\infty}^{+\infty}d\tau\,\del_\tau\left[\xi\left(\frac{1}{4e}\,\dot X^2-em^2\right)\right]\;.    
\end{equation}
For trajectories obeying $\dot X^\mu(\pm\infty)=0$, the action is gauge invariant provided that $\xi(+\infty)=\xi(-\infty)$ and $e(+\infty)=e(-\infty)$. We further demand that the einbein does not degenerate at infinity, which we ensure by imposing the asymptotic behavior $e(\pm\infty)=1$. This is the condition that makes the line physically infinite. Indeed, the asymptotic boundary condition on $e(\tau)$ implies that the integral $\int_{-\infty}^{+\infty} d\tau\,e(\tau)$ is divergent.

In this form, it is clear that there is no gauge invariant modulus for the einbein and one can gauge fix $e(\tau)=1$. Conversely, since the diffeomorphism parameter $\xi(\tau)$ does not have to vanish at the endpoints, the gauge fixed theory has a constant Killing vector for rigid translations. These features make the path integral quite different from the one of the finite line:
\begin{equation}
\int\frac{\cD X\cD e}{\rm VolGauge}\,e^{iS[X,e]}=\int \cD X\,e^{iS[X,1]}\int \cD c\cD b\,e^{iS_{\rm gh}} \;,   
\end{equation}
where the ghost action is the same as before. Due to the different boundary conditions, now the $c$ ghost admits a zero-mode, related to the rigid translation symmetry. When computing correlation functions, one can saturate the ghost zero-mode by inserting an ``unintegrated vertex operator'' $c(\tau_*)\,V_k(\tau_*)$ at a fixed time $\tau_*$. The ghost path integral can then be normalized to one again and discarded. The net difference from the case of the finite line can be summarized as follows
\begin{itemize}
\item One gauge fixes $e(\tau)=1$ and takes the range of the affine parameter\footnote{The same conclusion was reached in \cite{Mogull:2020sak} after explicitly performing the LSZ reduction.} $\tau\in\mathbb{R}$. There is no modulus for $e$. 
\item The gauge fixed theory admits a Killing vector for rigid translations. This leftover global symmetry can be used to fix the position of one vertex operator.
\item In the massive case, the mass term in \eqref{Infinite action} gives rise to an infinite phase. Upon regularization, this serves in implementing the LSZ reduction of the asymptotic states, which we will explain later on in this section.
\end{itemize}

The generating functional with vanishing Neumann conditions at infinity is essentially the same as \eqref{Z Neumann}. To obtain the one appropriate for the infinite line one first shifts $\tau\rightarrow\tau-\frac{T}{2}$ and then sends $T\rightarrow\infty$. Since this limit involves some subtleties, for now we keep $T$ fixed as an infrared regulator and
shift times in the Neumann propagator \eqref{G Neumann}, which becomes
\begin{equation}\label{G Neumann shifted}
G^{\rm shift}_{\rm NN}(\tau,\sigma)=|\tau-\sigma|-\frac{1}{T}(\tau^2+\sigma^2)-\frac{T}{6} \;.  
\end{equation}
In the limit $T\rightarrow\infty$ translation invariance is restored and the propagator reduces to $|\tau-\sigma|$ plus a divergent constant. Since the source of $Z_{\rm NN}[j]$ has no zero-mode, thanks to the delta function in \eqref{Z Neumann}, constant terms in the propagator do not contribute to the generating functional. The divergent term $-\frac{T}{6}$ thus drops from all correlation functions and we can safely use the propagator $|\tau-\sigma|$.
After removing the divergent part of the propagator, we can take the limit $T\rightarrow\infty$ yielding the path integral for the infinite line. Recalling that the trajectory is split as $X^\mu(\tau)=x_0^\mu+z^\mu(\tau)$, the path integral factorizes as the delta function for the zero-mode of the source times the normalized expectation value for $z^\mu$ as in \eqref{Z Neumann}:
\begin{equation}\label{Z infinite}
Z_\infty[j]:=\int \big[\cD X\big]_{\rm NN}\,e^{iS[X]+i\int_{-\infty}^{+\infty}d\tau\,j(\tau)\cdot X(\tau)}=(2\pi)^D\delta^D(j_0)\,\left\l e^{i\int_{-\infty}^{+\infty}d\tau\,j(\tau)\cdot z(\tau)}\right\r_{\infty} \;,    
\end{equation}
where the quadratic action $S[X]=\frac14\int_{-\infty}^{+\infty}d\tau\,\dot X^2$ is the gauge fixed version of \eqref{Infinite action} with the mass term removed. The normalized expectation value is given by
\begin{equation}\label{average infinity}
\left\l\exp\,i\int_{-\infty}^{+\infty}\!\!\!\!\!d\tau\,j_\mu(\tau)\,z^\mu(\tau)\right\r_{\infty}=\exp\left\{\frac{i}{2}\int_{-\infty}^{+\infty}\!\!\!\!\!d\tau d\sigma\,j^\mu(\tau)\,|\tau-\sigma|\,j_\mu(\sigma)\right\}\;,    
\end{equation}
corresponding to the two-point function $\big\l z^\mu(\tau)\,z^\nu(\sigma)\big\r_\infty=-i\,\eta^{\mu\nu}|\tau-\sigma|$.
Another subtlety with the limit $T\rightarrow\infty$ is related to the asymptotic conditions $\dot z^\mu(\pm\infty)=0$, because
the derivative of the translation-invariant propagator is the sign function $\epsilon(\tau-\sigma)$, which does not vanish for $\tau\rightarrow\pm\infty$. To recover the correct boundary conditions, one has to take the derivative of \eqref{G Neumann shifted} \emph{prior} to sending $T\rightarrow\infty$. Indeed, the derivative $\del_\tau G^{\rm shift}_{\rm NN}(\tau,\sigma)=\epsilon(\tau-\sigma)-\frac{2}{T}\,\tau$ does vanish for $\tau=\pm T/2$.

More importantly, the propagator itself diverges for large times. We will now study the scalar three-point function and see how this infrared divergence is responsible for implementing the LSZ reduction. To this end, we consider the path integral with three vertex operators
\begin{equation}\label{path test infinite}
\begin{split}
\int \big[\cD X\big]_{\rm NN}\,e^{ik_1\cdot X(-T/2)}e^{ik_2\cdot X(\tau_*)}e^{ik_3\cdot X(T/2)}\,e^{iS[X]}&=(2\pi)^D\delta^D(k_1+k_2+k_3)\\
&\times\left\l e^{ik_1\cdot z(-T/2)}e^{ik_2\cdot z(\tau_*)}e^{ik_3\cdot z(T/2)}\right\r_{\infty} \;.   
\end{split}
\end{equation}
Here we have taken all momenta to be incoming and we have regularized the asymptotic states by inserting them at times $\pm T/2$, to be sent to infinity at the end.
We compute the above correlator using \eqref{average infinity} with source $j^\mu(\tau)=\sum_ik_i^\mu\delta(\tau-\tau_i)$, yielding
\begin{equation}
\left\l e^{ik_1\cdot z(-T/2)}e^{ik_2\cdot z(\tau_*)}e^{ik_3\cdot z(T/2)}\right\r_{\infty}=\exp\Big\{-i\Big((T/2+\tau_*)k_1^2+(T/2-\tau_*)k_3^2\Big)\Big\}\;,    
\end{equation}
where we assumed momentum conservation, enforced by \eqref{path test infinite}. In the massive case, we regularize the infinite phase coming from \eqref{Infinite action} as $e^{-im^2T}$, so that
\begin{equation}
e^{-im^2T}\left\l e^{ik_1\cdot z(-T/2)}e^{ik_2\cdot z(\tau_*)}e^{ik_3\cdot z(T/2)}\right\r_{\infty}=e^{-i(T/2+\tau_*)(k_1^2+m^2)-i(T/2-\tau_*)(k_3^2+m^2)}\;.    
\end{equation}
Isolating the $T$-dependent part, one can see that in the limit $T\rightarrow\infty$ the correlator vanishes, unless the asymptotic particles $1$ and $3$ are on shell:
\begin{equation}
e^{-i\frac{T}{2}(k^2+m^2-i\epsilon)}\quad\xrightarrow{T\,\to\,\infty}\quad\left\{
\begin{array}{ll}
0 &{\rm if}\;k^2+m^2\neq0\\
1 &{\rm if}\;k^2+m^2=0
\end{array}
\right.  \;,  
\end{equation}
the same holding in the massless case.
We conclude that taking the limit $T\to\infty$ performs the LSZ reduction on the two legs corresponding to the asymptotic worldline states. If both particles $1$ and $3$ at asymptotic times are on shell, the correlator gives
\begin{equation}\label{A3 almost}
e^{-im^2T}\left\l e^{ik_1\cdot z(-T/2)}e^{ik_2\cdot z(\tau_*)}e^{ik_3\cdot z(T/2)}\right\r_{\infty}=1\;, 
\end{equation}
which has all legs amputated and does not depend on the regulator $T$.

Let us now show that, for the general case of an arbitrary number of vertex operators, the correlation function
\begin{equation}
e^{-im^2T}\Big\l\prod_{i=1}^Ne^{ik_i\cdot z(\tau_i)}\Big\r_\infty\;,
\end{equation}
does not depend on $T$ if the vertex operators creating the asymptotic states at the endpoints are on shell.
We choose the particles $1$ and $N$ to be the asymptotic states of the worldline, so we take $k_1^2=-m^2$, $\tau_1=-T/2$ and $k^2_N=-m^2$, $\tau_N=T/2$. The remaining $N-2$ momenta need not be on shell. The correlation function is evaluated as
\begin{equation}\label{correlator general}
e^{-im^2T}\Big\l\prod_{i=1}^Ne^{ik_i\cdot z(\tau_i)}\Big\r_\infty=\exp\Big\{i\sum_{i<j}k_i\cdot k_j\,|\tau_i-\tau_j|-im^2T\Big\}\;. 
\end{equation}
Using that $|T/2\pm\tau|=T/2\pm\tau$ for $T\rightarrow\infty$, the $T$-dependence drops out thanks to momentum conservation and the on-shell condition on $k_1$ and $k_N$. We can see this by isolating the $T$-dependent part of the sum in \eqref{correlator general}:
\begin{equation}
\begin{split}
\Big(\sum_{i<j}k_i\cdot k_j\,|\tau_i-\tau_j|-m^2T\Big)\Big\rvert_T&=\frac{T}{2}\,(k_1+k_N)\cdot\sum_{i=2}^{N-1}k_i+T\,k_1\cdot k_N-T\,m^2\\
&=-\frac{T}{2}\,(k_1+k_N)^2+T\,k_1\cdot k_N-T\,m^2=0\;.
\end{split}    
\end{equation}
In conclusion, we have shown that it is well-defined to create on-shell asymptotic states at the endpoints of the worldline by placing vertex operators at infinity. The propagators involving them are dealt with by the prescription of dropping any infinite time associated with the asymptotic states, as for instance in $|\infty\pm\tau|=\infty\pm\tau\rightarrow\pm\tau$.

If all external particles are on shell, 
path integrals with vertex operators on the infinite line compute certain contributions to scattering amplitudes in $\phi^3$ theory\footnote{In the next section we will generalize our construction to include full tree-level amplitudes in scalar theories with general polynomial interactions.}. As in \eqref{path test infinite}, the only role of the zero-mode of $X^\mu(\tau)$ is to generate the momentum-conserving delta function. For this reason, given a generic number of vertex operators  
\begin{equation}
V_i(\tau_i)=-ig\,e^{ik_i\cdot X(\tau_i)}\;,\quad X^\mu(\tau_i)=x_0^\mu+z^\mu(\tau_i)\;,\quad i=1,\ldots,N\;,    
\end{equation}
we denote their correlation function with the zero-mode removed by
\begin{equation}
\begin{split}
\Big\l\!\!\Big\l V_1(\tau_1)\cdots V_N(\tau_N)\Big\r\!\!\Big\r&:=e^{-ix_0\cdot(k_1+ \cdots +k_N)}\Big\l V_1(\tau_1)\cdots V_N(\tau_N)\Big\r_\infty\\
&=\Big\l e^{ik_1\cdot z(\tau_1)}\cdots e^{ik_N\cdot z(\tau_N)}\Big\r_\infty \;,   
\end{split}    
\end{equation}
so that the complete path integral with $N$ insertions factorizes as
\begin{equation}
\begin{split}
\int \big[\cD X\big]_{\rm NN}\,V_1(\tau_1)\cdots V_N(\tau_N)\,e^{iS[X]}&=
\int d^Dx_0\,\Big\l V_1(\tau_1)\cdots V_N(\tau_N)\Big\r_{\infty}\\
&=(2\pi)^D\delta^D(k_1+\cdots+k_N)\,\Big\l\!\!\Big\l V_1(\tau_1)\cdots V_N(\tau_N)\Big\r\!\!\Big\r\;.
\end{split}    
\end{equation}
When computing amplitudes, momentum conservation will always be assumed, but we will omit the delta function multiplying the normalized correlator. 
With this notation, the three-point amplitude is given directly by the correlator of three on-shell vertex operators. Two of them are inserted at plus and minus infinity, creating the asymptotic states at the endpoints, while we use translation invariance to fix the third at an arbitrary finite time, say $\tau=0$. The relevant correlator is \eqref{A3 almost} and
\begin{equation}
\label{3 point amplitude}
\cA_3=(-ig)^{-2}\Big\l\!\!\Big\l V_{1}(-\infty)\,V_{2}(0)\,V_{3}(+\infty)\Big\r\!\!\Big\r=-ig\;,   
\end{equation}
is the amplitude.
The prefactor of $(-ig)^{-2}$ removes the two extra powers of $-ig$ coming from $V_{1}(-\infty)$ and $V_{3}(+\infty)$, as these vertex operators create the asymptotic states and are not related to an actual cubic vertex. Similarly, the genus expansion of string theory assigns a prefactor of $g_s^{-2}$ to the path integral on the sphere.
A proper distinction between states and more general vertex operators will be discussed at length in the next section.

While the three-point amplitude is represented by a single correlator, higher point correlators capture only a certain class of Feynman diagrams in $\phi^3$ theory. For the general case of $N$ on-shell vertex operators, we choose the particles $1$ and $N$ to be at the worldline endpoints, with $\tau_1=-\infty$ and $\tau_N=+\infty$, and we use translation invariance to fix the position of $V_{2}(\tau_2)$ at $\tau_2=0$. The remaining $N-3$ vertex operators are integrated over the line. Let us mention that, in view of computing the full $N$-point amplitude, the choice of particles $1$, $2$ and $N$ does not matter, since the amplitude must be totally symmetric in the end. The integrated correlation function
\begin{equation}\label{Infinite ladder}
\begin{split}
\cA_N^{\rm ladder}&:=(-ig)^{-2}\prod_{k=3}^{N-1}\int_{-\infty}^{+\infty}\!\!\!d\tau_k\,\Big\l\!\!\Big\l V_{1}(-\infty)\,V_{2}(0)\,\prod_{i=3}^{N-1}V_{i}(\tau_i)\,V_{N}(+\infty)\Big\r\!\!\Big\r\\
&=(-ig)^{N-2}\prod_{k=3}^{N-1}\int_{-\infty}^{+\infty}\!\!\!d\tau_k\,\exp\Bigg\{\frac{i}{2}\sum_{i,j=1}^Nk_i\cdot k_j\,|\tau_i-\tau_j|\Bigg\}\;,   
\end{split}    
\end{equation}
gives the contribution to the $N$-point amplitude of all half-ladder diagrams, where the external particles $2,3,\ldots,N-1$ are attached to the line connecting particles $1$ and $N$. Diagrammatically, this can be visualized as 
\begin{center}
    \begin{tikzpicture}[scale=2]
        \node at (0,0) {$\cA^{\rm ladder}_N=$};
        \node at (1,0) {x};
        \node at (1,0.3) {$k_1$};
        \draw (1,0)--(5,0);
        \draw (2,0)--(2,1);
        \draw[snake it] (4,0)--(4,1);
        \node at (5,0) {x};
        \node at (1.7,1.1) {$k_2$};
        \node at (2.7,1.1) {$k_3$};
        \draw[snake it] (3,0)--(3,1);
        \node at (1,-0.5) {\small{$\dot{X}^\mu(-\infty)=0$}};
        \node at (5,-0.5) {\small{$\dot{X}^\mu(\infty)=0$}};
        \node at (2,-0.2) {\small{$0$}};
        \node at (3,-0.2) {\small{$\tau_3$}};
        \node at (4,-0.2) {\small{$\tau_{N-1}$}};
        \node at (4.3,1.1) {$k_{N-1}$};
        \node at (3.5,0.5) {$\dots$};
        \node at (5,0.3) {$k_N$};
    \end{tikzpicture}
\end{center}
As before, wavy lines denote integrated vertex operators, while the straight line at $\tau_2=0$ denotes the one fixed by translation invariance. The asymptotic endpoints of the worldline carry momenta $k_1$ and $k_N$, created by the vertex operators $V_{1}(-\infty)$ and $V_{N}(+\infty)$, respectively. These are depicted by a cross on the worldline endpoints.

Since the vertex operators are integrated over the entire line, \eqref{Infinite ladder} gives rise to $(N-2)!$ individual Feynman diagrams, depending on the relative positions of the vertex operators.
To appreciate this point, and see which contributions are missing, we compute explicitly the four-point case.
Using the aforementioned prescriptions to remove the infinite times, we obtain the correlator 
\begin{equation}
\begin{split}
\Big\l\!\!\Big\l V_{1}(-\infty)\,V_{2}(0)\,V_{3}(\tau)\,V_{4}(+\infty)\Big\r\!\!\Big\r&=(-ig)^4\exp\Big\{i\Big(\tau\,(k_1\cdot k_3-k_4\cdot k_3)+|\tau|\,k_2\cdot k_3\Big)\Big\}\\
&=(-ig)^4\left[\theta(\tau)\,e^{-i\tau\,(s+m^2)}+\theta(-\tau)\,e^{i\tau\,(u+m^2)}\right]\;,
\end{split}    
\end{equation}
where $\theta(\tau)$ is the Heaviside step function. In going to the second line we used momentum conservation, the on-shell condition $k_i^2+m^2=0$ and defined the Mandelstam variables
\begin{equation}
s=(k_1+k_2)^2\;,\quad
t=(k_2+k_3)^2\;,\quad
u=(k_1+k_3)^2\;.
\end{equation}
Integration over $\tau$ gives straightforwardly the $s$- and $u$-channels of the four-point amplitude:
\begin{equation}
\cA_4^{\rm ladder}=(-ig)^{-2}\int_{-\infty}^{+\infty}\!\!\!d\tau\,\Big\l\!\!\Big\l V_{1}(-\infty)\,V_{2}(0)\,V_{3}(\tau)\,V_{4}(+\infty)\Big\r\!\!\Big\r=\frac{ig^2}{s+m^2}+\frac{ig^2}{u+m^2}  \;, 
\end{equation}
which are the half-ladder contributions from the point of view of the line connecting particles 1 and 4. From this perspective, the missing $t$-channel is given by attaching the subtree with particles 2 and 3 to the main line connecting 1 and 4. 

In the general case \eqref{Infinite ladder}, we are missing all diagrams where the particles $2,3,\ldots,N-1$ are part of subtrees that then attach to the line connecting particles 1 and $N$.
In order to describe the full amplitude in a unified framework, we will introduce the idea of generalized vertex operators, which we will relate to the path integral on the semi-infinite line.

\subsection{Semi-infinite line: nonlinear fields and generalized vertex operators}\label{sec:semiinfinite}

For the sake of dealing with entire subtrees attached to the infinite worldline, we need to generalize the concept of vertex operator. Looking back at section \ref{sec:2.3}, the vertex operator for $\phi^3$ theory originated from the potential $-ig\,\phi(X)$ in the worldline action. More precisely, we can identify the $i$-th particle with momentum $k_i$ with the free field
\begin{equation}\label{phifree}
\phi_i(x):=e^{ik_i\cdot x}\;,\quad (\B-m^2)\phi_i=0\;,    
\end{equation}
obeying the Klein-Gordon equation. With this free field we then associate the on-shell vertex operator
\begin{equation}
\label{v1 plane}
V_i(\tau)=-ig\,\phi_i\big(X(\tau)\big)=-ig\,e^{ik_i\cdot X(\tau)}\;,    
\end{equation}
used in the previous subsection. 
In the case of the four-point amplitude discussed previously, including the missing $t$-channel requires to attach the subtree formed by particles 2 and 3 to the worldline. This subtree is described by the \emph{nonlinear field}
\begin{equation}\label{phi23}
\phi_{23}(x):=-\frac{g}{(k_2+k_3)^2+m^2}\,e^{i(k_2+k_3)\cdot x}\;.  
\end{equation}
In terms of standard Feynman rules, a factor of $-ig$ comes from the cubic vertex and another factor of $-i$ from the propagator. We call $\phi_{23}$ a nonlinear field in the sense that it obeys the nonlinear multiparticle  equation
\begin{equation}
(\B-m^2)\phi_{23}=g\,\phi_{2}\phi_{3}\;,   
\end{equation}
in terms of the linear fields \eqref{phifree}. We will discuss this point in detail in the next section.

In order to attach the subtree \eqref{phi23} to the worldline, we define the generalized vertex operator
\begin{equation}\label{V23}
V_{23}(\tau):=-ig\,\phi_{23}(X(\tau))=\frac{ig^2}{(k_2+k_3)^2+m^2}\,e^{i(k_2+k_3)\cdot X(\tau)}\;.   
\end{equation}
We can now insert $V_{23}$ into a three-point correlation function, together with the asymptotic states for particles 1 and 4. The general formula \eqref{correlator general} still applies, since only the momenta of the asymptotic states need to be on shell. We fix the position of $V_{23}$ at $\tau=0$ and obtain
\begin{equation}
(-ig)^{-2}\Big\l\!\!\Big\l V_{1}(-\infty)\,V_{23}(0)\,V_{4}(+\infty)\Big\r\!\!\Big\r=\frac{ig^2}{(k_2+k_3)^2+m^2}\;,
\end{equation}
which is exactly the contribution of the missing $t$-channel to the four-point amplitude. Altogether, the amplitude is given by
\begin{equation}
\label{4 point amplitude}
\cA_4=(-ig)^{-2}\Big\l\!\!\Big\l V_{1}(-\infty)\,\Big(\int_{-\infty}^{+\infty}\!\!\!d\tau\,V_{2}(0)\,V_{3}(\tau)+V_{23}(0)\Big)\,V_{4}(+\infty)\Big\r\!\!\Big\r  \;, 
\end{equation}
in terms of correlation functions on the infinite line. 
Graphically, we considered the following contributions:
\begin{center}
    \begin{tikzpicture}[scale=2]
    \node at (0.3,0) {$\cA_4=$};
    \node at (1,0) {x};
    \node at (1,0.3) {$k_4$};
    \draw (1,0)--(3,0);
    \draw (1.5,1)--(1.5,0);
    \node at (1.3,1) {$k_2$};
    \draw[snake it] (2.5,1)--(2.5,0);
    \node at (2.3,1) {$k_3$};
    \node at (1.5,-0.1) {\small{$0$}};
    \node at (2.5,-0.1) {\small{$\tau$}};
    \node at (3,0) {x};
    \node at (3,0.3) {$k_1$};
    
    \node at (3.5,0) {$+$};
    
    \node at (4,0) {x};
    \node at (4,0.3) {$k_4$};
    \draw (4,0)--(6,0);
    \draw (5,0.6)--(5,0);
    \draw (4.5,1)--(5,0.6)--(5.5,1);
    \node at (4.3,1) {$k_3$};
    \node at (5.7,1) {$k_2$};
    \node at (5,-0.1) {\small{$0$}};
    \node at (6,0) {x};
    \node at (6,0.3) {$k_1$};
\end{tikzpicture}
\end{center}
The first contribution comes from inserting two linear vertex operators in the infinite worldline, of which one is fixed at $\tau=0$ through translational invariance and the other one is integrated over. For the choice of worldline we have made, this gives the $s$ and $u$ channels. The second contribution comes from inserting one bilinear vertex operator, yielding the $t$ channel. Here we have chosen the worldline to connect particles 1 and 4, but we could have chosen any other combination. The full four-point amplitude would still be given by two correlators, one involving a bilinear vertex $V_{ij}$. In the next section we will generalize the formula \eqref{4 point amplitude} to all orders.

Although it looks like $V_{23}$ and $\phi_{23}$ are the same thing, they are conceptually different objects. $\phi_{23}(x)$ is a field, described by a tree diagram with root at the spacetime point $x^\mu$. The operation mapping $\phi_{23}(x)\longmapsto V_{23}(\tau)$ is the act of gluing this rooted tree to the worldline. The fact that this gluing just amounts to the pullback of $x^\mu$ to the line and adding a factor of $-ig$ is an accident of $\phi^3$ theory. In the next section we will see that this distinction is necessary when dealing with different interactions. Even in the case of $\phi^3$ interactions, the distinction between the linear field $\phi_i(x)$ and vertex operator $-ig\,\phi_i(X(\tau))$
explains the prefactor of $(-ig)^{-2}$ in front of all correlation functions: the operators creating the asymptotic states at infinity should be thought of as the pullback of
fields $\phi_{i}(X(\pm\infty))$, rather than vertex operators.

\subsubsection*{The Dirichlet-Neumann path integral}

The introduction of generalized vertex operators, related to nonlinear multiparticle fields, allows one to represent full tree-level amplitudes as correlation functions on the infinite line. To complete the picture, we need to discuss how to compute these multiparticle fields. In the remainder of this section, we will argue that these are also given by worldline path integrals, albeit with a different geometry: the semi-infinite line. 

We use the term semi-infinite line for a worldline of infinite proper length (in the same sense as discussed in sec.~\ref{sec:3.2}), but with a fixed boundary. More precisely, we will consider trajectories parametrized by $\tau\in[0,+\infty)$ where $X^\mu(0)=x^\mu$ is a fixed spacetime point, while at infinity we will create a momentum eigenstate $\ket{k}$ by acting with a vertex operator on the zero-momentum state $\ket{0}$. As we have discussed in the previous subsection, the projection onto the zero-momentum state corresponds to the boundary condition $\dot X^\mu(+\infty)=0$. To determine the path integral for this geometry, we start from the reparametrization invariant action
\begin{equation}\label{Semi-Infinite action}
S[X,e]=\int_{0}^{\infty}d\tau\,\left[\,\frac{1}{4e}\,\dot X^2-em^2\right]\;,    
\end{equation}
with mixed Dirichlet-Neumann boundary conditions
\begin{equation}
X^\mu(0)=x^\mu\;,\quad \dot X^\mu(+\infty)=0\;.
\end{equation}
As in sec.~\ref{sec:3.2} we impose the asymptotic behavior $e(+\infty)=1$, which implies that the proper length $\int_0^\infty d\tau\,e(\tau)$ is infinite. Correspondingly, there is no gauge invariant modulus for $e(\tau)$. The boundary condition at $\tau=0$ now requires that the diffeomorphism parameter obeys $\xi(0)=0$. This does not allow for a constant Killing vector after gauge fixing, as it is clear from the presence of the boundary at $\tau=0$, which breaks translation invariance. These properties are reflected by the ghosts: neither $c$ nor $b$ admit a zero-mode and the ghost path integral can be factored out. We summarize here the prescriptions for the path integral on the semi-infinite line: 
\begin{itemize}
\item One gauge fixes $e(\tau)=1$ and takes the range of the affine parameter $\tau\in[0,+\infty)$. There is no modulus for $e$. 
\item The gauge fixed theory has no Killing vector for rigid translations. All vertex operators (except the one creating the asymptotic state at infinity) must be integrated over the worldline.
\item In the massive case, the mass term in \eqref{Semi-Infinite action} yields again an infinite phase. Regularized with a large propagation time $T$, it contributes to the LSZ reduction on the asymptotic state.
\end{itemize}

We are now ready to compute the generating functional. To this end, we split the trajectory with Dirichlet-Neumann boundary conditions as
\begin{equation}\label{DN BC}
X^\mu(\tau)=x^\mu+z^\mu(\tau)\;,\quad z^\mu(0)=0\;,\quad\dot z^\mu(+\infty)=0\;,    
\end{equation}
where $x^\mu$ is the fixed spacetime point implementing the boundary condition at $\tau=0$. We denote the path integral measure by $\big[\cD X\big]_x\equiv\big[\cD z\big]_{\rm DN}$, where the subscript DN refers to the boundary conditions \eqref{DN BC}.
This path integral computes correlators between a position eigenstate fixed at $x^\mu$ and the zero-momentum eigenstate. As before, we create the eigenstate of momentum $k^\mu$ by acting with the vertex operator $e^{ik\cdot X(+\infty)}$ at the asymptotic time. We denote the generating functional by
\begin{equation}\label{Z semiinfinite}
Z_{\rm DN}[j]\equiv\prescript{}{x}{\!\!\left\l\exp\,i\int_{0}^{\infty}d\tau\,j_\mu(\tau)\,X^\mu(\tau)\right\r_{\rm DN}}:=\int \big[\cD z\big]_{\rm DN} \,e^{i\int_{0}^{\infty}d\tau\,\left(\frac14\,\dot X^2+j_\mu X^\mu\right)} \;,
\end{equation}
to keep track of the dependence on the fixed position $X^\mu(0)=x^\mu$. Since the action does not depend on it, $x^\mu$ only appears in a plane wave with the zero-mode of the current:
\begin{equation}
\prescript{}{x}{\!\!\left\l\exp\,i\int_{0}^{\infty}d\tau\,j_\mu(\tau)\,X^\mu(\tau)\right\r_{\rm DN}}=e^{ij_{0}\cdot x}\exp\left\{\frac{i}{2}\int_{0}^{\infty}d\tau d\sigma\,\Big(j^\mu(\tau)\,G_{\rm DN}(\tau,\sigma)\,j_\mu(\sigma)\Big)\right\}\;.   
\end{equation}
The propagator is uniquely determined by the boundary conditions $G_{\rm DN}(0,\sigma)=0$ and
\\
$\del_\tau G_{\rm DN}(+\infty,\sigma)=0$, yielding
\begin{equation}\label{G DN big}
\begin{split}
G_{\rm DN}(\tau,\sigma)&=|\tau-\sigma|-(\tau+
\sigma)\;,\\
\del_\tau G_{\rm DN}(\tau,\sigma)&=\epsilon(\tau-\sigma)-1\;. 
\end{split}    
\end{equation}
We can check that this is the correct generating functional by comparing a correlation function with the corresponding transition amplitude in the operator formulation. We thus compute the expectation value of two vertex operators, one inserted at time $\tau$ and one at infinity, which we regulate by a large time $T$:
\begin{equation}\label{DN test}
\begin{split}
\prescript{}{x}{\!\!\Big\l e^{ip\cdot X(\tau)}e^{ik\cdot X(T)}\Big\r_{\rm DN}}&=e^{i(k+p)\cdot x}\exp\Big\{\frac{i}{2}\,\Big(k^2\,G_{\rm DN}(T,T)+p^2\,G_{\rm DN}(\tau,\tau)+2\,k\cdot p\,G_{\rm DN}(\tau,T)\Big)\Big\}\\
&=e^{i(k+p)\cdot x}\exp\Big\{-i\Big(k^2\,(T-\tau)+(k+p)^2\,\tau\Big)\Big\}\;,   \end{split}    
\end{equation}
upon using the propagator \eqref{G DN big}. First of all, we see that the plane wave carries the total momentum, which will always be the case for correlators of vertex operators.
We can now check that \eqref{DN test} reproduces the operator transition amplitude:
\begin{equation}
\begin{split}
\bra{-k}e^{-i(T-\tau)\hat P^2}e^{ip\cdot \hat X}e^{-i\tau \hat P^2}\ket{x}&=e^{-i(T-\tau)k^2}     \bra{-k}e^{ip\cdot \hat X}e^{-i\tau \hat P^2}\ket{x}\\
&=e^{-i(T-\tau)k^2}     \bra{-k-p}e^{-i\tau \hat P^2}\ket{x}\\
&=e^{-i(T-\tau)k^2}e^{-i\tau(k+p)^2} e^{i(k+p)\cdot x}\;.
\end{split}
\end{equation}
As before, one can distribute masses by adding the global factor $e^{-im^2T}$, resulting in
\begin{equation}\label{DN test 2}
e^{-im^2T}\prescript{}{x}{\!\!\Big\l e^{ip\cdot X(\tau)}e^{ik\cdot X(T)}\Big\r_{\rm DN}}
=e^{i(k+p)\cdot x}\exp\Big\{-i\Big((k^2+m^2)\,(T-\tau)+\big((k+p)^2+m^2\big)\,\tau\Big)\Big\}\;.   
\end{equation}
At this point, taking the limit $T\rightarrow\infty$ makes the correlator vanish, unless the state at the asymptotic time is on shell, exactly as in the case of the infinite line. The difference, of course, is that here the LSZ reduction is performed on the single asymptotic state. In a similar way, one proves
that the infrared regulator $T$ drops from all correlation functions, provided that the state at infinity is on shell.

We can now use this result to show that the path integral over the semi-infinite line generates the nonlinear field \eqref{phi23}. We create the on-shell state of momentum $k_3$ by inserting the corresponding vertex operator at infinity, while the vertex operator for the particle 2 is integrated over the worldline. With the normalization \eqref{v1 plane} we obtain
\begin{equation}\label{phi23 PI}
\begin{split}
(-ig)^{-1}\int_0^\infty\!\!\! d\tau\prescript{}{x}{\!\!\Big\l V_2(\tau)V_3(+\infty)\Big\r_{\rm DN}}&=(-ig)\int_0^\infty\!\!\! d\tau\, e^{i(k_2+k_3)\cdot x}e^{-i\tau\,((k_2+k_3)^2+m^2)} \\
&=-\frac{g}{(k_2+k_3)^2+m^2}\,e^{i(k_2+k_3)\cdot x}\;,
\end{split}    
\end{equation}
which is exactly the field $\phi_{23}(x)$ in \eqref{phi23}. Analogous to the infinite line, the prefactor $(-ig)^{-1}$ removes the extra factor of the coupling coming from the vertex operator at infinity. Interestingly, in string theory the path integral on the disk is weighted by $g_s^{-1}$.
In the next section we will generalize these examples to all orders, showing that the infinite and semi-infinite lines generate all possible tree diagrams.

\section{Recursive relations for trees}\label{sec:recursions}

In this section, we will consider nonlinear multiparticle fields of arbitrary order, which are given by all rooted tree diagrams. We will show that they are recursively generated by the path integral on the semi-infinite line, and can be used to define generalized vertex operators. Inserting these into an infinite worldline yields arbitrary $N$-point tree-level amplitudes as worldline correlators. Finally, we will generalize the results for $\phi^3$ theory to scalar theories with arbitrary polynomial interactions. For brevity, throughout this section we will consider massless fields, but the massive case can be obtained by just changing the on-shell condition on the external momenta.

\subsection{Multiparticle fields from perturbative solutions}\label{sec:Perturbsolution}

In the previous section we gave an example of how composite fields, such as $\phi_{23}(x)$ in \eqref{phi23}, can be represented as Dirichlet-Neumann worldline path integrals. Before proceeding with this perspective, we take a brief detour to review how multiparticle fields of any order are extracted from perturbative solutions of the nonlinear field equations \cite{Boulware:1968zz,Monteiro:2011pc,Nutzi:2018vkl,Bonezzi:2023xhn}. This field theoretic derivation will be instrumental in section \ref{sec:cutandglue} to prove worldline recursive relations.

We start with massless $\phi^3$ theory, with action $S[\phi]=\int d^Dx\left(-\frac12\,(\del\phi)^2-\frac{g}{3!}\,\phi^3\right)$. In order to solve the field equation $\B \phi=\frac{g}{2}\,\phi^2$  perturbatively, one makes the following ansatz for the field:
\begin{equation}\label{phi ansatz}
\phi(x)=\sum_{n=1}^\infty \frac{1}{n!}\,\phi^{(n)}(x)\;.
\end{equation}
Plugging the ansatz into the field equation and matching both sides, order by order in the total \emph{arity}\footnote{This is effectively a power series in the coupling constant $g$ since, applying the recursion, $\phi^{(n)}$ scales as $g^{n-1}$.} $n$, at lowest order one finds that $\phi^{(1)}(x)$ is a free field obeying $\B\phi^{(1)}=0$. To higher orders, $\phi^{(n)}$ is determined as an $n$-linear function of $\phi^{(1)}$ via the recursive relation
\begin{equation}
\label{nonlinear phi3 eom}
\phi^{(n)}=\frac{1}{\B}\,\frac{g}{2}\sum_{k+l=n}\binom{n}{k}\phi^{(k)}\phi^{(l)}=\frac{1}{2}\sum_{k+l=n}\binom{n}{k}\quad\vcenter{\hbox{\begin{tikzpicture}
        \filldraw (5,0) circle (1pt);
        \node at (5,-0.3) {$\scriptstyle x$};
        \draw (5,0)--(6,0);
        \draw (6,0)--(7,1);
        \draw (6,0)--(7,-1);
        \filldraw[gray!60] (7,1) circle (0.5); 
        \filldraw[gray!60] (7,-1) circle (0.5); 
        \node at (7,1) {$\phi^{(k)}$};
        \node at (7,-1) {$\phi^{(l)}$};
    \end{tikzpicture}}}\quad,
\end{equation}
where $\frac{1}{\B}$ is a shorthand notation for the convolution with the Feynman propagator in position space.
The recursive formula \eqref{nonlinear phi3 eom} generates $\frac{1}{n!}\phi^{(n)}(x)$ as the sum of all unlabeled cubic trees, with root at the spacetime point $x^\mu$ and $n$ external legs. By unlabeled diagrams we mean that all external legs (here given by factors of $\phi^{(1)}$) are identical. In the expression for $\frac{1}{n!}\phi^{(n)}(x)$, each diagram $\Gamma$ contributes with a weight $\frac{1}{S_\Gamma}$, where $S_\Gamma$ is its symmetry factor.
This weight emerges as follows: at each step of the recursion, all the newly generated diagrams are distinct. For each individual diagram, the symmetry factor is built order by order from the twofold symmetry of the rooted cubic vertex. As an explicit example, the first few orders resulting from \eqref{nonlinear phi3 eom} are given by 
\begin{equation}
\begin{split}
\frac{1}{2!}\,\phi^{(2)}&=\frac{1}{2}\;\vcenter{\hbox{\begin{tikzpicture}
        \filldraw (0,0) circle (1pt);
        \node at (0,-0.3) {$\scriptstyle x$};
        \draw (0,0)--(1,0);
        \draw (1,0)--(2,0.8);
        \draw (1,0)--(2,-0.8);
    \end{tikzpicture}}} \;,\quad 
\frac{1}{3!}\,\phi^{(3)}=\frac{1}{2}\;\vcenter{\hbox{\begin{tikzpicture}
        \filldraw (0,0) circle (1pt);
        \node at (0,-0.3) {$\scriptstyle x$};
        \draw (0,0)--(1,0);
        \draw (1.5,0.5)--(2,0);
        \draw (1,0)--(2,1);
        \draw (1,0)--(2,-1);
    \end{tikzpicture}}} \quad,\\
\frac{1}{4!}\,\phi^{(4)}&=\frac{1}{2}\;\vcenter{\hbox{\begin{tikzpicture}
        \filldraw (0,0) circle (1pt);
        \node at (0,-0.3) {$\scriptstyle x$};
        \draw (0,0)--(1,0);
        \draw (1.7,0.7)--(2,0.4);
        \draw (1.4,0.4)--(2,-0.2);
        \draw (1,0)--(2,1);
        \draw (1,0)--(2,-1);
    \end{tikzpicture}}}\;+\frac{1}{2^3}\;\vcenter{\hbox{\begin{tikzpicture}
        \filldraw (0,0) circle (1pt);
        \node at (0,-0.3) {$\scriptstyle x$};
        \draw (0,0)--(1,0);
        \draw (1.6,0.6)--(2,0.2);
        \draw (1.6,-0.6)--(2,-0.2);
        \draw (1,0)--(2,1);
        \draw (1,0)--(2,-1);
    \end{tikzpicture}}} \quad,   
\end{split}
\end{equation}
in terms of ordinary Feynman diagrams, where the external legs are factors of $\phi^{(1)}$, internal lines are propagators $\frac{i}{\B}$ and vertices carry a factor of $-ig$. 

The recursive formula \eqref{nonlinear phi3 eom} gives the fields $\phi^{(n)}$ as $n$-linear functions of the single argument $\phi^{(1)}$, so that we refer to them as \emph{unpolarized} nonlinear fields. The multiparticle fields we are interested in, on the other hand, are described by rooted trees where the external legs are distinct on-shell particles, i.e. momentum eigenstates with momenta $k_i^\mu$, obeying $k_i^2=0$. Since labeled trees (meaning all external legs are distinct) have a trivial symmetry factor, each diagram contributes to a multiparticle field with unit weight.
In general, a multiparticle field of order $n$ depends $n$-linearly on the particle labels $\{i_1,\ldots,i_n\}$, which form a word $I\equiv i_1i_2\dots i_n$ of length $|I|=n$.
In the following, we will denote these multiparticle fields as $\phi_{I}(x)$. Since the $n$ particles $i_1,i_2,\ldots,i_n$ are assigned to the external legs in all possible ways, $\phi_I$ is totally symmetric in the labels and one needs to consider only \emph{unordered} words.

Given an unpolarized nonlinear field of order $n$, one can extract the multiparticle field $\phi_{12\cdots n}$ by \emph{polarizing} its expression as a function of $\phi^{(1)}$. To this end, one starts from $\frac{1}{n!}\,\phi^{(n)}$ as obtained from \eqref{nonlinear phi3 eom} and finds $\phi_{12\cdots n}$ by applying the following procedure:
\begin{itemize}
\item In all external legs, specialize the free field $\phi^{(1)}$ to the sum of on-shell momentum eigenstates
\begin{equation*}
\phi^{(1)}(x)=\phi_1(x)+\phi_2(x)+\cdots+\phi_n(x)\;,\quad \phi_i(x)=e^{ik_i\cdot x} \;.
\end{equation*}
\item Keep only the terms multilinear in $12\cdots n$, with no repeated letters.
\end{itemize}
This procedure sums over all permutations of $1,2,\ldots,n$ and thus distributes the particle labels on the external legs in all possible ways, ensuring total symmetry. Moreover, the sum over the $S_\Gamma$ permutations that leave a given diagram $\Gamma$ invariant cancels the weight $\frac{1}{S_\Gamma}$. In this way, all labeled trees contribute to $\phi_{12\cdots n}$ with weight one, as expected. To clarify this procedure, we apply it to polarize the trilinear field $\phi^{(3)}$. One starts with the substitution
\begin{equation}
\begin{split}
\frac{1}{3!}\,\phi^{(3)}&=\frac{1}{2}\;\vcenter{\hbox{\begin{tikzpicture}
        \filldraw (0,0) circle (1pt);
        \node at (0,-0.3) {$\scriptstyle x$};
        \node at (2.4,1) {$\scriptstyle \phi^{(1)}$};
        \node at (2.4,-1) {$\scriptstyle \phi^{(1)}$};
        \node at (2.4,0) {$\scriptstyle \phi^{(1)}$};
        \draw (0,0)--(1,0);
        \draw (1.5,0.5)--(2,0);
        \draw (1,0)--(2,1);
        \draw (1,0)--(2,-1);
    \end{tikzpicture}}}\quad\longrightarrow\quad
    \phi_{123}=\frac{1}{2}\;\left.\vcenter{\hbox{\begin{tikzpicture}
        \filldraw (0,0) circle (1pt);
        \node at (0,-0.3) {$\scriptstyle x$};
        \node at (2.6,1) {$\scriptstyle 1+2+3$};
        \node at (2.6,-1) {$\scriptstyle 1+2+3$};
        \node at (2.6,0) {$\scriptstyle 1+2+3$};
        \draw (0,0)--(1,0);
        \draw (1.5,0.5)--(2,0);
        \draw (1,0)--(2,1);
        \draw (1,0)--(2,-1);
    \end{tikzpicture}}}\right\rvert_{123}\quad,
\end{split}    
\end{equation}
where ${}\rvert_{123}$ instructs to keep only the trilinear part with no repeated labels. Summing over permutations, and using the twofold symmetry of the upper subtree, finally yields
\begin{equation}\label{phi123 QFT}
\phi_{123}=\vcenter{\hbox{\begin{tikzpicture}
        \filldraw (0,0) circle (1pt);
        \node at (0,-0.3) {$\scriptstyle x$};
        \node at (2.2,1) {$\scriptstyle 1$};
        \node at (2.2,0) {$\scriptstyle 2$};
        \node at (2.2,-1) {$\scriptstyle 3$};
        \draw (0,0)--(1,0);
        \draw (1.5,0.5)--(2,0);
        \draw (1,0)--(2,1);
        \draw (1,0)--(2,-1);
    \end{tikzpicture}}}\;+\;\vcenter{\hbox{\begin{tikzpicture}
        \filldraw (0,0) circle (1pt);
        \node at (0,-0.3) {$\scriptstyle x$};
        \node at (2.2,1) {$\scriptstyle 2$};
        \node at (2.2,0) {$\scriptstyle 3$};
        \node at (2.2,-1) {$\scriptstyle 1$};
        \draw (0,0)--(1,0);
        \draw (1.5,0.5)--(2,0);
        \draw (1,0)--(2,1);
        \draw (1,0)--(2,-1);
    \end{tikzpicture}}}\;+\;\vcenter{\hbox{\begin{tikzpicture}
        \filldraw (0,0) circle (1pt);
        \node at (0,-0.3) {$\scriptstyle x$};
        \node at (2.2,1) {$\scriptstyle 3$};
        \node at (2.2,0) {$\scriptstyle 1$};
        \node at (2.2,-1) {$\scriptstyle 2$};
        \draw (0,0)--(1,0);
        \draw (1.5,0.5)--(2,0);
        \draw (1,0)--(2,1);
        \draw (1,0)--(2,-1);
    \end{tikzpicture}}}  \quad,
\end{equation}
correctly displaying all three channels and unit weight for all diagrams. In the next subsection we will resume with the worldline representation of multiparticle fields and scattering amplitudes.

\subsection{Multiparticle fields and amplitudes as worldline correlators}

In section \ref{sec:semiinfinite} we have argued that multiparticle fields can be obtained as worldline correlation functions with Dirichlet-Neumann boundary conditions. We also introduced multilinear vertex operators, related to multiparticle fields by e.g. $V_{23}(\tau)=-ig\,\phi_{23}(X(\tau))$.
We will now demonstrate how nonlinear fields of all orders can be computed recursively, by inserting generalized vertex operators of lower order on the semi-infinite worldline. 

The starting point of the recursive procedure is a linear field $\phi_i(x)=e^{ik_i\cdot x}$, corresponding to a single external particle. Consistently, it can also be represented by a Dirichlet-Neumann worldline, on which we create a single state of momentum $k_i$ at the asymptotic boundary:
\begin{equation}\label{linear phi DN}
\begin{tikzpicture}[scale=1.8]
    \node at (0.5,0) {$\phi_i(x)=$};
    \filldraw (1,0) circle (1pt);
    \draw (1,0)--(3,0);
    \node at (3,0) {x};
    \node at (1,-0.3) {$\scriptstyle X(0)=x$};
    \node at (3,-0.3) {$\scriptstyle\dot{X}(\infty)=0$};
    \node at (3,0.3) {$k_i$};
    \node at (4.5,0) {$=\prescript{}{x}{\!\Big\langle e^{ik_i\cdot X(\infty)}\Big\rangle_{\rm DN}}= e^{ik_i\cdot x}\;.$};\,
\end{tikzpicture}    
\end{equation}
We now take a moment to clarify our notation. The diagram depicts a semi-infinite worldline, in which the dotted endpoint is the fixed one with Dirichlet boundary conditions, while the crossed one denotes that we have created an asymptotic on-shell state of momentum $k_i$ at infinity. Given a rooted tree, we will refer to the choice of worldline as the choice of which particle is interpreted as the asymptotic state. 
As mentioned in section \ref{sec:semiinfinite}, the linear field can be glued to another worldline by mapping it to the linear vertex operator $V_i(\tau)=-ig\phi_i(X(\tau))$.
The bilinear field $\phi_{ij}(x)$ can be then derived from inserting a $V_i(\tau)$ on the worldline describing $\phi_j(x)$, or vice versa. The vertex operator $V_i(\tau)$ has to be integrated, due to the lack of translation invariance. We represent this by the following diagram, recalling that a wavy line stands for an integrated vertex operator
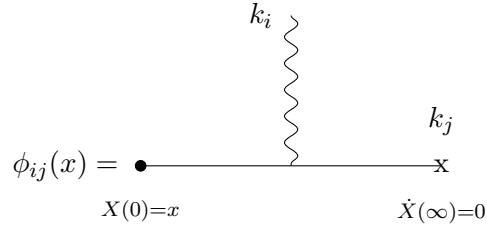
\begin{center}
    \begin{tikzpicture}[scale=2]
    \node at (0.5,0) {$\phi_{ij}(x)=$};
    \filldraw (1,0) circle (1pt);
    \draw (1,0)--(3,0);
    \draw[snake it] (2,1)--(2,0);
    \node at (1.8,1) {$k_i$};
    \node at (3,0) {x};
    \node at (1,-0.3) {$\scriptstyle X(0)=x$};
    \node at (3,-0.3) {$\scriptstyle\dot{X}(\infty)=0$};
    \node at (3,0.3) {$k_j$};
\end{tikzpicture}
\end{center}
The analytic expression for the above diagram is the massless version of \eqref{phi23 PI}
\begin{equation}
    \phi_{ij}(x)=\int_0^\infty d\tau\prescript{}{x}{\!\Big\langle  V_i(\tau)\phi_j(X(\infty))\Big\rangle_{\rm DN}}={-g}\,\frac{e^{ik_{ij}\cdot x}}{k_{ij}^2},
\end{equation}
where we denote the sum of momenta by $k_{i_1\cdots i_n}:=k_{i_1}+\cdots+k_{i_n}$. Let us stress that the state at infinity is created by the pullback of the linear field, differing by a factor $-ig$ from the vertex operator. This distinction will be crucial later on.
Here, the choice of asymptotic state does not affect the result for the diagram, as $\phi_{ij}$ is evidently symmetric in the labels. 
Once again, one can map this field to a vertex operator $V_{ij}(\tau)=-ig\phi_{ij}(X(\tau))$ in order to insert it in some other worldline. 
In fact, this is exactly the vertex operator used to compute the $t$-channel of the four-point amplitude in \eqref{4 point amplitude}, by inserting it in an infinite Neumann-Neumann worldline.

Moving forward to the three-particle field, we choose particle 1 to be the state at infinity. To add the remaining two particles, one needs to take into account all possible ways to create them: we can either insert two linear vertex operators or a bilinear one. Summing over the two possibilities yields
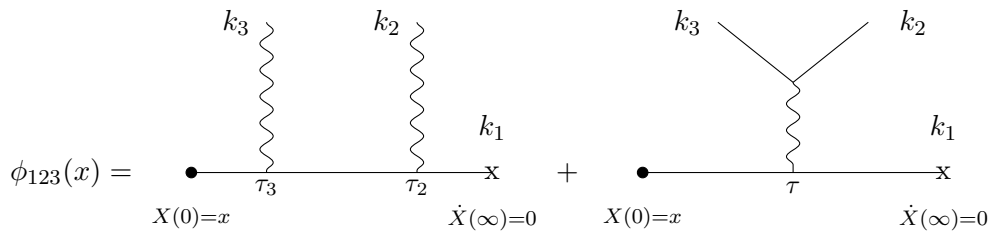
\begin{center}
    \begin{tikzpicture}[scale=2]
    \node at (0.2,0) {$\phi_{123}(x)=$};
    \filldraw (1,0) circle (1pt);
    \draw (1,0)--(3,0);
    \draw[snake it] (1.5,1)--(1.5,0);
    \node at (1.3,1) {$k_3$};
    \draw[snake it] (2.5,1)--(2.5,0);
    \node at (2.3,1) {$k_2$};
    \node at (1.5,-0.1) {\small{$\tau_3$}};
    \node at (2.5,-0.1) {\small{$\tau_2$}};
    \node at (3,0) {x};
    \node at (1,-0.3) {$\scriptstyle X(0)=x$};
    \node at (3,-0.3) {$\scriptstyle\dot{X}(\infty)=0$};
    \node at (3,0.3) {$k_1$};
    
    \node at (3.5,0) {$+$};
    
    \filldraw (4,0) circle (1pt);
    \draw (4,0)--(6,0);
    \draw[snake it] (5,0.6)--(5,0);
    \draw (4.5,1)--(5,0.6)--(5.5,1);
    \node at (4.3,1) {$k_3$};
    \node at (5.8,1) {$k_2$};
    \node at (5,-0.1) {\small{$\tau$}};
    \node at (6,0) {x};
    \node at (4,-0.3) {$\scriptstyle X(0)=x$};
    \node at (6,-0.3) {$\scriptstyle\dot{X}(\infty)=0$};
    \node at (6,0.3) {$k_1$};
\end{tikzpicture}
\end{center}
Notice that the first contribution generates two distinct Feynman diagrams, depending on the relative position of the vertex operators for particles 2 and 3, as it can be seen from the explicit expression
\begin{align}\label{phi123}
     \phi_{123}(x)&=\int_0^\infty\!\!\! d\tau_2\int_0^\infty\!\!\! d\tau_3\prescript{}{x}{\!\Big\langle V_2(\tau_2)V_3(\tau_3)\phi_1(X(\infty))\Big\rangle_{\rm DN}}+\int_0^\infty\!\!\! d\tau\prescript{}{x}{\!\Big\langle V_{23}(\tau)\phi_1(X(\infty))\Big\rangle_{\rm DN}} \notag \\
     &=g^2\,\frac{e^{ik_{123}\cdot x}}{k_{123}^2}\left(\frac{1}{k^2_{12}}+\frac{1}{k^2_{13}}+\frac{1}{k^2_{23}}\right)\;.
 \end{align}
Although the starting point in \eqref{phi123} is only manifestly symmetric in the exchange of particles 2 and 3, by adding both contributions the final result is totally symmetric in the three particle labels and coincides with \eqref{phi123 QFT}. This shows that a different choice of worldline would give the same result, with a reshuffling of the channels produced by each correlator.

In order to find an emerging pattern, we give also the next two orders of multiparticle fields, only in terms of diagrams: 
\begin{equation}\label{phi1234}
\begin{tikzpicture}
    \node at (-0.5,0) {$\phi_{1234}(x)=$};
    \filldraw (2,0) circle (1pt);
    \draw (2,0)--(5,0);
    \draw[snake it] (2.5,1)--(2.5,0);
    \node at (2.2,1) {$4$};
    \draw[snake it] (3.5,1)--(3.5,0);
    \node at (3.2,1) {$3$};
    \draw[snake it] (4.5,1)--(4.5,0);
    \node at (4.2,1) {$2$};
    \node at (5,0) {x};
    \node at (5.2,0.3) {$1$};

    \node at (6,0) {$+$};

    \filldraw (7,0) circle (1pt);
    \draw (7,0)--(10,0);
    \draw[snake it] (8,0.5)--(8,0);
    \draw (7.5,1)--(8,0.5)--(8.5,1);
    \node at (7.3,1.1) {$4$};
    \node at (8.3,1.1) {$3$};
    \draw[snake it] (9,1)--(9,0);
    \node at (9.2,1.1) {$2$};
    \node at (10,0) {x};
    \node at (10.2,0.3) {$1$};
    
\node at (11,0) {$+$};
    \node at (12,0) {2 more};

\node at (1.5,-2) {$+$};
    \filldraw (2,-2) circle (1pt);
    \draw (2,-2)--(5,-2);
    \draw[snake it] (3.5,-2)--(3.5,-1.5);
    \filldraw[color=gray!60] (3.5,-1) circle (0.5);
    \node at (3.5,-1) {$432$};
    \node at (5,-2) {x};
    \node at (5.2,-1.7) {$1$};
    
\end{tikzpicture}
\end{equation}

\begin{equation}
\begin{tikzpicture}
    \node at (-0.5,0) {$\phi_{1...5}(x)=$};
    \filldraw (1,0) circle (1pt);
    \draw (1,0)--(5,0);
    \draw[snake it] (1.5,1)--(1.5,0);
    \node at (1.2,1) {$5$};
    \draw[snake it] (2.5,1)--(2.5,0);
    \node at (2.2,1) {$4$};
    \draw[snake it] (3.5,1)--(3.5,0);
    \node at (3.2,1) {$3$};
    \draw[snake it] (4.5,1)--(4.5,0);
    \node at (4.2,1) {$2$};
    \node at (5,0) {x};
    \node at (5.2,0.3) {$1$};

    \node at (6,0) {$+$};

    \filldraw (7,0) circle (1pt);
    \draw (7,0)--(11,0);
    \draw[snake it] (8,0.5)--(8,0);
    \draw (7.5,1)--(8,0.5)--(8.5,1);
    \node at (7.3,1.1) {$5$};
    \node at (8.3,1.1) {$4$};
    \draw[snake it] (9,1)--(9,0);
    \node at (9.2,1.1) {$3$};
    \draw[snake it] (10,1)--(10,0);
    \node at (10.2,1.1) {$2$};
    \node at (11,0) {x};
    \node at (11.2,0.3) {$1$};
    \node at (12.5,0) {+ 5 more};

    \node at (0.3,-2) {$+$};
    \filldraw (1,-2) circle (1pt);
    \draw (1,-2)--(5,-2);
    \draw[snake it] (2,-1.5)--(2,-2);
    \draw (1.5,-1)--(2,-1.5)--(2.5,-1);
    \node at (1.2,-1) {$5$};
    \node at (2.2,-1) {$4$};
    \draw[snake it] (4,-1.5)--(4,-2);
    \draw (3.5,-1)--(4,-1.5)--(4.5,-1);
    \node at (3.2,-1) {$3$};
    \node at (4.2,-1) {$2$};
    \node at (5,-2) {x};
    \node at (5.2,-1.7) {$1$}; 

    \node at (6.5,-2){+ 2 more +};

    \filldraw (8,-2) circle (1pt);
    \draw (8,-2)--(11,-2);
    \draw[snake it] (9,-1)--(9,-2);
    \node at (9,-0.8) {$5$};
    \draw[snake it] (10,-1.5)--(10,-2);
    \node at (11,-2) {x};
    \node at (11.2,-1.7) {$1$};
    \filldraw[color=gray!60] (10,-1) circle (0.5);
    \node at (10,-1) {$432$};
    \node at (12.5,-2){+ 3 more};

    \node at (0.3,-4) {$+$};
    \filldraw (1,-4) circle (1pt);
    \draw (1,-4)--(5,-4);
    \draw[snake it] (3,-4)--(3,-3.5);
    \filldraw[color=gray!60] (3,-3) circle (0.6);
    \node at (3,-3) {$5432$};
    \node at (5,-4) {x};
    \node at (5.2,-3.7) {$1$};
    
\end{tikzpicture}
\end{equation}
where we fixed particle 1 to be the state at infinity.
The gray circles with words of particle labels, e.g. $I=432$, denote the insertion of the generalized vertex operator $V_{I}(\tau)=-ig\,\phi_I(X(\tau))$. Here, by ``more'' we mean diagrams that differ from the previous one only by a permutation of the particle labels $2,3,\ldots,N$, since particle 1 is always fixed as the boundary state of the worldline. To sum over inequivalent permutations, we make use of the already established symmetry of the previous orders. Namely, $n$-linear vertex operators and fields are totally symmetric in the arguments and are therefore labeled by unordered words of length $n$. Moreover, there is an explicit symmetry in permuting the words associated with different vertex operators, since they are all integrated over the line. This is the reason why a single worldline correlator with $m$ insertions generates $m!$ distinct Feynman diagrams.

To summarize, multiparticle fields can be derived by inserting vertex operators of lower order into the semi-infinite worldline, with one chosen particle as the state at infinity. These generalized vertex operators are related to the fields as their pullback to the worldline, multiplied by the coupling constant:
\begin{equation}
    V_{I}(\tau):= -ig\,\phi_{I}(X(\tau))\;.
\end{equation}
To make expressions more compact, we will often denote an integrated vertex operator with a bar, such as
\begin{equation}
\overline{V}_I:=\int_0^\infty d\tau\,V_I(\tau)\;.
\end{equation}
The few explicit orders presented above suggest a general recursive structure. To make it more transparent, we will now unpolarize the multiparticle fields and associated vertex operators.

\subsubsection*{All order recursion for nonlinear fields}\label{sec:recursion}

To unpolarize the multiparticle fields, we identify all external legs by removing their labels. Analytically, we do so by substituting all linear fields $\phi_i(x)=e^{ik_i\cdot x}$, describing individual external particles, with a generic free field $\phi^{(1)}(x)$ obeying the massless Klein-Gordon equation:
\begin{equation}
\phi_i(x)\quad\longrightarrow\quad \phi^{(1)}(x)\;,\quad \B\phi^{(1)}=0\;.    
\end{equation}
This is the opposite procedure to the one described in section \ref{sec:Perturbsolution}. Under this identification, all multiparticle fields $\phi_I(x)$ of a given length $|I|=n$ reduce to the nonlinear field $\phi^{(n)}(x)$ introduced there. We thus define the corresponding unpolarized vertex operator and its integrated version as
\begin{equation}\label{phi n and v n}
V^{(n)}(\tau)=-ig\,\phi^{(n)}(X(\tau))\;,\quad \overline{V}^{(n)}=\int_0^\infty d\tau\,V^{(n)}(\tau)\;.    
\end{equation}

Applying this procedure to the examples displayed above, we obtain the following expressions for the nonlinear fields $\phi^{(3)}$, $\phi^{(4)}$ and $\phi^{(5)}$:
\begin{equation}\label{nonlinear phi345}
\begin{split}
\frac{1}{2!}\,\phi^{(3)}(x)&=\prescript{}{x}{\!\!\left\langle\left[\frac{1}{2}\left(\overline{V}^{(1)} \right)^2+\frac{1}{2}\overline{V}^{(2)} \right]\phi^{(1)}(X(\infty))\right\rangle_{\rm DN}}\;,\\[2mm]
\frac{1}{3!}\,\phi^{(4)}(x)&=\prescript{}{x}{\!\!\left\langle\left[\frac{1}{3!}\left(\overline{V}^{(1)} \right)^3+\frac{1}{2}\overline{V}^{(2)}\overline{V}^{(1)}+\frac{1}{3!}\overline{V}^{(3)} \right]\phi^{(1)}(X(\infty))\right\rangle_{\rm DN}}\;,\\[2mm]
\frac{1}{4!}\,\phi^{(5)}(x)&=\prescript{}{x}{\!\!\left\langle\left[\frac{1}{4!}\left(\overline{V}^{(1)} \right)^4+\frac{1}{2^2}\left(\overline{V}^{(1)} \right)^2\overline{V}^{(2)}+\frac{1}{3!}\overline{V}^{(3)}\overline{V}^{(1)}\right.\right.}\\
&\hspace{29mm}+\left.\left.\frac{1}{2^3}\left(\overline{V}^{(2)} \right)^2+\frac{1}{4!}\overline{V}^{(4)} \right]\phi^{(1)}(X(\infty))\right\rangle_{\rm DN}\;,
\end{split}    
\end{equation}
In the unpolarized form \eqref{nonlinear phi345}, the nonlinear fields display a clear recursive pattern that leads us to propose the all-order ansatz
\begin{equation}\label{relation with exponential}
\sum_{m=1}^\infty \frac{1}{(m-1)!}\phi^{(m)}(x)=\prescript{}{x}{\!\!\left\langle\exp\left(\sum_{k=1}^\infty \frac{1}{k!}\overline{V}^{(k)}\right)\phi^{(1)}(X(\infty))\right\rangle_{\rm DN}}\;,
\end{equation}
which is meant to be matched order by order in the total arity $m$. This is our central result. One may notice that the different symmetry factor on the left-hand side, compared to \eqref{phi ansatz}, is due to singling out one linear field at infinity. In section \ref{sec:cutandglue} we will prove that the nonlinear fields obtained from \eqref{relation with exponential} coincide with the perturbative solutions \eqref{nonlinear phi3 eom}. 

We shall now explain how to polarize \eqref{relation with exponential} to extract the multiparticle fields with distinct external particles. This is slightly different from the procedure outlined in section \ref{sec:Perturbsolution}, since the above recursive relation singles out the worldline asymptotic endpoint. More precisely, given the expression for $\frac{1}{(n-1)!}\phi^{(n)}(x)$ from \eqref{relation with exponential}, one finds the multiparticle field $\phi_{12\cdots n}(x)$ by applying the following steps: 
\begin{itemize}
\item Replace the free field at infinity by $$\phi^{(1)}(X(\infty))\;\longrightarrow\; \phi_1(X(\infty))\equiv e^{ik_1\cdot X(\infty)}\;.$$ This amounts to choosing particle 1 to be the asymptotic state of the worldline.
\item Substitute the unpolarized vertex operators $\overline{V}^{(k)}$ with
\begin{equation*}
\overline{V}^{(k)}\;\longrightarrow\;\overline{V}_{\underbrace{\scriptstyle(2+3+\cdots n)(2+3+\cdots n)\cdots(2+3+\cdots n)}_{k\,{\rm times}}} \;,   
\end{equation*}
where the formal sum of words is defined by linearity, e.g. $\overline{V}_{(i_1+i_2)j}:=\overline{V}_{i_1j}+\overline{V}_{i_2j}$. This corresponds to taking the free field in the vertex operators to be the sum of plane waves describing particles $2,3,\ldots,n$, i.e. $\phi^{(1)}(x)=\phi_2(x)+\cdots+\phi_n(x)$ .
\item Use total symmetry of the multilinear vertex operators $\overline{V}_{i_1i_2\cdots i_k}$. For the final result, keep only the terms with all distinct letters $23\cdots n$.
\end{itemize}
To exemplify this procedure, we apply it to obtain $\phi_{1234}(x)$ from the nonlinear field $\phi^{(4)}(x)$. Starting from the expression
\begin{equation}
\frac{1}{3!}\,\phi^{(4)}(x)=\prescript{}{x}{\!\!\left\langle\left[\frac{1}{3!}\left(\overline{V}^{(1)} \right)^3+\frac{1}{2}\overline{V}^{(2)}\overline{V}^{(1)}+\frac{1}{3!}\overline{V}^{(3)} \right]\phi^{(1)}(X(\infty))\right\rangle_{\rm DN}}\;, 
\end{equation}
as it comes from \eqref{relation with exponential}, we follow the steps outlined above and obtain
\begin{equation}
\begin{split}
\phi_{1234}(x)&=\prescript{}{x}{\!\!\left\langle\left[\frac{1}{3!}\Big(\overline{V}_{2+3+4} \Big)^3+\frac{1}{2}\,\overline{V}_{(2+3+4)(2+3+4)}\,\overline{V}_{2+3+4}\right.\right.}\\
&\hspace{33mm}+\left.\left.\frac{1}{3!}\,\overline{V}_{(2+3+4)(2+3+4)(2+3+4)} \right]_{234}\phi_1
(X(\infty))\right\rangle_{\rm DN}\\[2mm]
&=\prescript{}{x}{\!\!\Big\langle\Big[\overline{V}_2\overline{V}_3\overline{V}_4+\overline{V}_{23}\,\overline{V}_{4}+\overline{V}_{42}\,\overline{V}_{3}+\overline{V}_{34}\,\overline{V}_{2}+\overline{V}_{234} \Big]\phi_1
(X(\infty))\Big\rangle_{\rm DN}}\;,
\end{split}    
\end{equation}
which is the analytic expression corresponding to \eqref{phi1234}. One sees that $\phi_{1234}(x)$ is indeed given by the sum of all rooted tree diagrams\footnote{Recall that $m$ integrated vertex operators generate $m!$ different Feynman diagrams, depending on the relative order of their insertions. The five correlators above correspond to 15 distinct Feynman diagrams, upon recalling that $\overline{V}_{234}$ alone has three diagrams in it.} with external particles $1234$, each diagram contributing with unit weight.

\subsubsection*{Tree-level scattering amplitudes}

We conclude this part by giving our prescription for computing scattering amplitudes.
Given the generalized vertex operators $V_I(\tau)$ up to order $|I|=N-2$, one can obtain the $N$-point scattering amplitude by inserting them into an infinite Neumann-Neumann worldline. For scattering of particles $1,2,\ldots,N$ we choose the infinite worldline to connect particles 1 and $N$. These are created at the asymptotic boundaries by the linear fields $\phi_1(X(-\infty))$ and $\phi_N(X(+\infty))$, respectively. The remaining $N-2$ particles are inserted into the worldline via generalized vertex operators of all possible arities, from a maximum number of $N-2$ insertions of linear vertex operators, to a single insertion of the $(N-2)$-linear operator $V_{23\cdots N-1}(\tau)$. Translation symmetry allows us to fix the position of one vertex operator, which we place at $\tau=0$. This leads to the following representation for the $N$-point scattering amplitude
\begin{align}
    \cA_N=\Big\langle\!\!\Big\langle \phi_1(X(-\infty)) \sum_{m=1}^{N-2}\sum_{I_1I_2\cdots I_{m}=23\cdots N-1}\hspace{-8mm} V_{I_1}(0)\prod_{k=2}^m\int_{-\infty}^{+\infty}\!\!\!\!\!\! d\tau_k\, V_{I_k}(\tau_k)\,\phi_N(X(+\infty))\Big\rangle\!\!\Big\rangle\;,
\end{align}
where $\sum_{I_1I_2\cdots I_{m}=23\cdots N-1}$ denotes the sum over all possible unordered partitions of $23\cdots N-1$ into $m$ unordered words. For instance, for the 5-point amplitude the relevant partitions of 234 are $2|3|4$, $23|4$, $24|3$, $34|2$, $234$. The corresponding 5 correlation functions give rise to the 15 diagrams contributing to the amplitude.

\subsection{Cutting and gluing worldlines}\label{sec:cutandglue}

In this subsection, we will prove that the worldline recursive formula \eqref{relation with exponential} yields the correct nonlinear fields, by showing that it is equivalent to the field theoretic recursion \eqref{nonlinear phi3 eom}. While the latter singles out the first cubic vertex from the root, \eqref{relation with exponential} singles out one external leg and is thus nontrivial to prove that they are equivalent.
To this end, in the following we will manipulate the right-hand side of \eqref{relation with exponential} into a form which displays the field theoretic cubic vertex as the gluing of three worldlines.

Schematically, we can expose a cubic vertex from \eqref{relation with exponential} by cutting the path integral open at the insertion point of a generalized vertex operator. In doing so, the path integral factorizes into three separate worldlines: the first, from the fixed boundary of the original line to the insertion point, is a finite Dirichlet-Dirichlet worldline. The second one arises as the detached vertex operator at the cutting point, while the third one is the remaining semi-infinite worldline, with fixed boundary at the cutting point and Neumann conditions at infinity. The three lines are then glued together by integrating over the spacetime position of the cutting point, thus giving a worldline representation of the cubic vertex. To make this idea more precise, we will now discuss in some detail how to cut and glue worldline path integrals in different ways. 

Given a quantum mechanical path integral, with some operator insertions, the concept of cutting it open is realized by adding an intermediate boundary condition to the trajectories. By integrating over this extra boundary condition, one recovers the original functional integral
as the gluing of two separate ones. The simplest example in this sense is given by a path integral with Dirichlet boundary conditions $X^\mu(0)=x^\mu$ and $X^\mu(T)=y^\mu$. One can decompose the functional space of trajectories by further fixing $X^\mu(\tau_*)=z^\mu$, where $\tau_*\in[0,T]$ is an intermediate time, and then integrating over $z^\mu$. Without operator insertions, this is just the statement that the quantum mechanical transition amplitude $K(x,y;T)=\bra{y}e^{-i\hat HT}\ket{x}$ obeys the group property
\begin{equation}
\int d^D z\,K(x,z;T_1)\,K(z,y;T_2)=K(x,y;T_1+T_2)\;. 
\end{equation}

The situation is more subtle in the case of relativistic worldlines. In this case, the propagation times and operator positions are moduli to be integrated over and one has to take care of possible changes in the moduli spaces, under the cutting or gluing procedure. For the open worldlines discussed so far the moduli spaces differ, depending on the three geometries analyzed in section \ref{sec:wline geometries}:
\begin{itemize}
\item The finite Dirichlet-Dirichlet line, with two fixed boundaries and $N$ vertex operator insertions, has $N+1$ moduli: the positions $\tau_i\in[0,T]$ of all vertex operators, as well as the proper length of the line $T$. The modulus $T$ has to be integrated from zero to infinity, while all vertex operators are integrated from zero to $T$.
\item The semi-infinite Dirichlet-Neumann line, with one fixed boundary, one momentum eigenstate at infinity and $N$ more vertex operator insertions, has $N$ moduli: the positions $\tau_i\in[0,+\infty)$ of all vertex operators, except the one at the asymptotic endpoint.
\item The infinite Neumann-Neumann line, with two momentum eigenstates at plus and minus infinity and $N$ more vertex operator insertions, has $N-1$ moduli: the \emph{relative} positions $\tau_i-\tau_j\in\mathbb{R}$ of the vertex operators, except the ones at the asymptotic endpoints. One usually uses translation invariance to fix the position of one operator and integrates over the positions of the remaining ones.
\end{itemize}

In all these cases, the cutting procedure can be performed by adding a Dirichlet boundary at an intermediate time: $X^\mu(\tau_*)=z^\mu$, and integrating over $z^\mu$. Whether the number of moduli is preserved, compared to the standard one described above, depends on the position of the cut: if the worldline is cut into two parts at a free propagation point, the glued path integral has one modulus less compared to the split ones. This is because gluing two edges of lengths $T_1$ and $T_2$ into a smooth edge of length $T\equiv T_1+T_2$ reduces the number of moduli by one, as after the gluing only the total length $T$ matters. 
If the cut, instead, is at the position of an operator insertion, the number of moduli is preserved, since the two lengths $T_1$ and $T_2$ are both relevant to determine the position of the insertion. 

To elucidate this procedure, let us work it out for the simplest example: the bilinear field $\phi^{(2)}(x)$ which, according to \eqref{relation with exponential}, is given by
\begin{equation}\label{phi2 correlator}
\phi^{(2)}(x)=\int_0^\infty \!\!dT\prescript{}{x}{\!\Big\langle V^{(1)}(T)\,\phi^{(1)}
(X(\infty))\Big\rangle_{\rm DN}}\;,
\end{equation}
where we denoted the insertion time by capital $T$, not to confuse it with the affine parameter of the semi-infinite worldline. We now rewrite \eqref{phi2 correlator} by making the path integral explicit, as
\begin{equation}
\phi^{(2)}(x)=\int_0^\infty \!\!dT\int \big[\cD X\big]_{X(0)=x}^{\dot X(\infty)=0}\,V^{(1)}(T)\,\phi^{(1)}
(X(\infty))\,e^{i\int_{0}^{\infty}d\tau\,\left(\frac14\,\dot X^2\right)}  \;,  
\end{equation}
where we indicated the full boundary conditions on the measure. To cut and glue the path integral at the insertion point $\tau=T$, we fix the trajectory as $X^\mu(T)=y^\mu$ and then integrate over $y^\mu$. This is just a rewriting of the same functional space, with the measure factorizing as
\begin{equation}
\big[\cD X\big]_{X(0)=x}^{\dot X(\infty)=0}=\int d^Dy\,\big[\cD X\big]_{X(0)=x}^{X(T)=y}\,\big[\cD X\big]_{X(T)=y}^{\dot X(\infty)=0} \;. 
\end{equation}
In the path integral, the exponential of the action similarly factorizes \begin{equation}
e^{i\int_{0}^{\infty}d\tau\,\left(\frac14\,\dot X^2\right)}=e^{i\int_{0}^{T}d\tau\,\left(\frac14\,\dot X^2\right)}e^{i\int_{T}^{\infty}d\sigma\,\left(\frac14\,\dot X^2\right)} \;,  
\end{equation} 
and the vertex operator at $\tau=T$ is ``detached'' from the functional integral, thanks to the intermediate boundary condition at $T$:
\begin{equation}
V^{(1)}(T)=-ig\,\phi^{(1)}(X(T))\equiv -ig\,\phi^{(1)}(y)\;.    
\end{equation}
The full path integral describing $\phi^{(2)}(x)$ thus splits as a free Dirichlet-Dirichlet line, times a Dirichlet-Neumann one and the field $\phi^{(1)}(y)$, all glued at the spacetime point $y$ as
\begin{equation}\label{cut example1}
\begin{split}
\phi^{(2)}(x)&=-ig\int d^Dy\,\phi^{(1)}(y)\int_0^\infty \!\!dT\\
&\left(\int \big[\cD X\big]_{X(0)=x}^{ X(T)=y}\,e^{i\int_{0}^{T}d\tau\,\left(\frac14\,\dot X^2\right)}\right)\times\left(\int \big[\cD X\big]_{X(T)=y}^{\dot X(\infty)=0}\phi^{(1)}
(X(\infty))\,e^{i\int_{T}^{\infty}d\tau\,\left(\frac14\,\dot X^2\right)}\right) \;.
\end{split}    
\end{equation}
Note that, even though the second worldline is parametrized from $T$ to infinity, its path integral does not depend on $T$: one can shift the affine parameter as $\tau=\tau'+T$ and rename the functional integration variable $X(\tau)=X'(\tau')$, to obtain the standard Dirichlet-Neumann correlator
\begin{equation}
\begin{split}
\int \big[\cD X\big]_{X(T)=y}^{\dot X(\infty)=0}\phi^{(1)}
(X(\infty))\,e^{i\int_{T}^{\infty}d\tau\,\left(\frac14\,\dot X^2\right)}&=\int \big[\cD X\big]_{X(0)=y}^{\dot X(\infty)=0}\phi^{(1)}
(X(\infty))\,e^{i\int_{0}^{\infty}d\tau\,\left(\frac14\,\dot X^2\right)}\\
&=\prescript{}{y}{\!\!\Big\langle \phi^{(1)}
(X(\infty))\Big\rangle_{\rm DN}}=\phi^{(1)}(y)\;,
\end{split}
\end{equation}
where we removed all primes after the shift and in the last equality we have recognized the ``identity correlator'' \eqref{linear phi DN} for a free on-shell field\footnote{To see this for a generic field obeying $\B\phi^{(1)}=0$ it is sufficient to express $\phi^{(1)}$ as an on-shell Fourier transform.}. Using this in \eqref{cut example1} and recalling the definition \eqref{free dd heat kernel} of the free partition function, we find the following form of $\phi^{(2)}$, completely factorized:
\begin{equation}\label{cut example2}
\begin{split}
\phi^{(2)}(x)
&=-ig\int d^Dy\,\int_0^\infty \!\!dT\,K_0(x,y;T)\,\phi^{(1)}(y)\prescript{}{y}{\!\!\Big\langle \phi^{(1)}
(X(\infty))\Big\rangle_{\rm DN}}\\
&=-ig\int d^Dy\,Z_0(x,y)\,\phi^{(1)}(y)\,\phi^{(1)}(y)\;.
\end{split}    
\end{equation}
Notice that the modulus $T$, originally associated with the position of the vertex operator, is interpreted after the cutting as the length of the Dirichlet-Dirichlet worldline, showing that the number of moduli is indeed preserved.

Since the free partition function is nothing but the field theory propagator
\begin{equation}\label{Z propagator}
Z_0(x,y)=-i\int\frac{d^Dp}{(2\pi)^D}\,\frac{e^{ip\cdot(x-y)}}{p^2-i\epsilon}\;, \quad \B_xZ_0(x,y)=i\,\delta^D(x-y)\;,   
\end{equation}
c.f. \eqref{dirichlet dirichlet path integral}, the above rewriting of $\phi^{(2)}$ has made contact with the standard Feynman rules, where the cubic vertex is expressed as an integral over the common boundary of three worldlines.
We can depict this cutting and gluing process as follows:
\begin{center}
    \begin{tikzpicture}[scale=1.5]
        \node at (0.2,0) {$\phi^{(2)}(x)=$};
        \node at (1,-0.3) {$\scriptstyle X(0)=x$};
        \filldraw (1,0) circle (1pt);
        \draw(1,0)--(3,0);
        \draw[snake it] (2,0)--(2,1);
        \node at (3,0) {x};
        \node at (3,-0.3) {$\scriptstyle \dot{X}(\infty)=0$};

        \node at (3.4,0) {$\equiv$};
        \node at (3.8,0) {$\Big{(}$};
        \filldraw (4,0) circle (1pt);
        \draw (4,0)--(5,0);
        \filldraw (5,0) circle (1pt);
        \node at (5.2,0) {$\Big{)}$};
        \node at (4,-0.3) { $\scriptstyle X(0)=x$};
        \node at (5,-0.3) {$\scriptstyle X(T)=y$};

        \node at (5.5,0) {$\sqcup$};
        
        \node at (6,0) {$-ig\Big{(}$};
        \filldraw (6.5,0) circle (1pt);
        \node at (6.5,-0.2) {$\scriptstyle y$};
        \draw (6.5,0)--(6.5,1);
        \node at (6.5,1) {x};
        \node at (6.8,0) {$\Big{)}$};

\node at (7.05,0) {$\sqcup$};
        \node at (7.3,0) {$\Big{(}$};
        \filldraw (7.5,0) circle (1pt);
        \draw (7.5,0)--(8.5,0);
        \node at (8.5,0) {x};
        \node at (8.7,0) {$\Big{)}$};
        \node at (7.5,-0.3) {$\scriptstyle X(T)=y$};
        \node at (8.5,-0.3) {$\scriptstyle \dot{X}(\infty)=0$};
        
    \end{tikzpicture}
\end{center}
with the cup symbols indicating the gluing by integration over the common boundary point $y^\mu$. In the form \eqref{cut example2}, the nonlinear field $\phi^{(2)}$ is manifestly the same as the one obtained from \eqref{nonlinear phi3 eom}. Applying the massless Klein-Gordon operator removes the propagator $Z_0$ (see \eqref{Z propagator}), showing that $\phi^{(2)}$ is the order $g$ solution to the $\phi^3$ field equation, i.e.
\begin{equation}
\B\phi^{(2)}(x)=g\,\phi^{(1)}(x)\phi^{(1)}(x)\;.    
\end{equation}

Before delving into the general proof, we shall give a second example of cutting the path integral. We consider a Dirichlet-Neumann correlator with two insertions of integrated vertex operators, of arbitrary arity $l$ and $m$:
\begin{equation}\label{cut example VlVm}
\prescript{}{x}{\!\Big\langle \overline{V}^{(l)}\,\overline{V}^{(m)}\,\phi^{(1)}
(X(\infty))\Big\rangle_{\rm DN}}=\int_0^\infty\!\!\!d\tau_1\int_0^\infty\!\!\!d\tau_2\prescript{}{x}{\!\Big\langle{V}^{(l)}(\tau_1)\,{V}^{(m)}(\tau_2)\,\phi^{(1)}
(X(\infty))\Big\rangle_{\rm DN}}   \;.     
\end{equation}
In order to cut the path integral at the position of the earliest operator, one has to first split the integration ranges into definite time orderings, since both operators are integrated over the whole line, namely
\begin{equation}\label{integration range split}
\int_0^\infty\!\!\!d\tau_1\int_0^\infty\!\!\!d\tau_2\,{V}^{(l)}(\tau_1)\,{V}^{(m)}(\tau_2)=\int_0^\infty\!\!\!dT\int_T^\infty\!\!\!d\tau_*\,{V}^{(l)}(T)\,{V}^{(m)}(\tau_*)+(l\leftrightarrow m)\;,    
\end{equation}
where $T$ denotes the time of the earliest insertion. We now add the boundary condition $X^\mu(T)=y^\mu$ and integrate over $y$. As before, in doing so the vertex operator at time $T$ detaches, e.g. $V^{(l)}(T)=-ig\,\phi^{(l)}(y)$. The path integral between $X^\mu(0)=x^\mu$ and $X^\mu(T)=y^\mu$ is again a free Dirichlet-Dirichlet line with modulus $T$, while the remaining Dirichlet-Neumann correlator is defined on the line with affine parameter $\tau\in[T,\infty)$. Similar to the previous case, this correlator does not depend on $T$, as one can shift the affine parameter $\tau=\tau'+T$, as well as the insertion time $\tau_*=\tau_*'+T$ and redefine $X(\tau)=X'(\tau')$ in the path integral. This leads to identify
\begin{equation}
\begin{split}
\int_T^\infty\!\!\! d\tau_* \big[\cD X\big]_{X(T)=y}^{\dot X(\infty)=0}V^{(m)}(\tau_*)\,\phi^{(1)}
(X(\infty))\,e^{i\int_{T}^{\infty}d\tau\,\left(\frac14\,\dot X^2\right)}&=\int_0^\infty\!\!\!d\tau_*\prescript{}{y}{\!\Big\langle{V}^{(m)}(\tau_*)\,\phi^{(1)}
(X(\infty))\Big\rangle_{\rm DN}}\\
&=\prescript{}{y}{\!\Big\langle\overline{V}^{(m)}\phi^{(1)}
(X(\infty))\Big\rangle_{\rm DN}}\;,
\end{split}  
\end{equation}
as a standard Dirichlet-Neumann correlator in the interval $[0,\infty)$. Applying this cutting and gluing procedure to \eqref{cut example VlVm} we end up with the factorized form
\begin{equation}
\prescript{}{x}{\!\Big\langle \overline{V}^{(l)}\,\overline{V}^{(m)}\,\phi^{(1)}
(X(\infty))\Big\rangle_{\rm DN}}=-ig\int d^Dy\,Z_0(x,y)\,\phi^{(l)}(y)\prescript{}{y}{\!\Big\langle \overline{V}^{(m)}\phi^{(1)}
(X(\infty))\Big\rangle_{\rm DN}} +(l\leftrightarrow m)\;,   
\end{equation}
which exhibits the free propagator and the first cubic vertex at the gluing point $y^\mu$:
\begin{center}
    \begin{tikzpicture}[scale=1.5]
        \filldraw (0,0) circle (1pt);
        \node at (0,-0.3) {$\scriptstyle X(0)=x$};
        \draw(0,0)--(3,0);
        \draw[snake it] (1,0)--(1,1);
        \filldraw[gray!60] (1,1) circle (0.35);
        \node at (1,1) {$V^{(l)}$};
        \draw[snake it] (2,0)--(2,1);
        \filldraw[gray!60] (2,1) circle (0.35);
        \node at (2,1) {$V^{(m)}$};
        \node at (3,0) {x};
        \node at (2.8,-0.3) {$\scriptstyle \dot{X}(\infty)=0$};

        \node at (3.4,0) {$\equiv$};
        \node at (3.8,0) {$\Big{(}$};
        \filldraw (4,0) circle (1pt);
        \draw (4,0)--(5,0);
        \filldraw (5,0) circle (1pt);
        \node at (5.2,0) {$\Big{)}$};
        \node at (4,-0.4) { $\scriptstyle X(0)=x$};
        \node at (5,-0.4) {$\scriptstyle X(T)=y$};

        \node at (5.5,0) {$\sqcup$};

        \node at (6,0) {$-ig\Big{(}$};
        \filldraw (6.5,0) circle (1pt);
        \node at (6.5,-0.4) {$\scriptstyle y$};
        \draw (6.5,0)--(6.5,1);
        \filldraw[gray!60] (6.5,1) circle (0.35);
        \node at (6.5,1) {$\phi^{(l)}$};
        \node at (6.8,0) {$\Big{)}$};

\node at (7.05,0) {$\sqcup$};
        \node at (7.3,0) {$\Big{(}$};
        \filldraw (7.5,0) circle (1pt);
        \draw (7.5,0)--(8.5,0);
        \draw[snake it] (8,0)--(8,1);
        \filldraw[gray!60] (8,1) circle (0.35);
        \node at (8,1) {$V^{(m)}$};
        \node at (8.5,0) {x};
        \node at (8.7,0) {$\Big{)}$};
        \node at (7.5,-0.4) {$\scriptstyle X(T)=y$};
        \node at (8.5,-0.4) {$\scriptstyle \dot{X}(\infty)=0$};

        \node at (9.5,0) {$+(l\leftrightarrow m)$};
    \end{tikzpicture}
\end{center}

\subsubsection*{Proof of the recursive relation}

We can now put to fruition the cutting and gluing procedure, to demonstrate that \eqref{relation with exponential} results in the same nonlinear fields as \eqref{nonlinear phi3 eom} to all orders. To this end, we will prove that $\phi^{(n)}$ in \eqref{relation with exponential} obeys
\begin{equation}\label{phi eom binomial}
\B\phi^{(n)}=\frac{g}{2}\sum_{k+l=n}\binom{n}{k}\phi^{(k)}\phi^{(l)}\;, 
\end{equation}
for all $n\geq2$ and is thus the order $g^{n-1}$ solution to the field equations. We start from the expression for $\phi^{(N+1)}(x)$, with arbitrary $N\geq1$. From the exponential formula \eqref{relation with exponential}, this is given by isolating  the terms where the integrated vertex operators count for a total arity of $N$:
\begin{align}\label{phi N+1}
    \phi^{(N+1)}(x)&=
    N!\sum_{l=1}^{N}\frac{1}{l!}\sum_{m_1+\dots+m_l=N}\prescript{}{x}{\!\Big\langle}\prod_{i=1}^l  \frac{1}{m_i!}\overline{V}^{(m_i)}\phi^{(1)}(X(\infty))\Big\rangle_{\rm DN}\;.
\end{align}
Each separate term has $l$ insertions of integrated vertex operators. $l$ ranges from 1, corresponding to a single insertion of $\overline{V}^{(N)}$, to $N$, where all vertex operators are linear.

In order to cut the path integral at the position of the earliest operator, we have to split the integration ranges as in \eqref{integration range split}. For $l$ vertex operator insertions, we have $l$ possible cases, in which the first operator is $V^{(m_i)}(\tau_i)$ for $i=1,2,\ldots,l$, respectively. Since \eqref{phi N+1} is totally symmetric in the labels $m_i$, all cases are equivalent, resulting in an overall factor of $l$:
\begin{equation}
\begin{split}
\sum_{m_1+\dots+m_l=N}\prod_{i=1}^l  \frac{1}{m_i!}\overline{V}^{(m_i)}&=\sum_{m_1+\dots+m_l=N}\prod_{i=1}^l\frac{1}{m_i!}\int_0^\infty\!\!\!d\tau_i\,{V}^{(m_i)}(\tau_i)\\
&=l\hspace{-5mm}\sum_{r+m_1+\dots+m_{l-1}=N}\frac{1}{r!}\int_0^\infty\!\!\!dT\,V^{(r)}(T)\prod_{i=1}^{l-1}\frac{1}{m_i!}\int_T^\infty\!\!\!d\tau_i\,{V}^{(m_i)}(\tau_i)\;,
\end{split}
\end{equation}
where the earliest insertion time has been denoted by $T$, and the order of the corresponding vertex operator by $r$. 
We can now cut and glue the worldline at time $\tau=T$ by fixing $X^\mu(T)=y^\mu$, while simultaneously integrating over $y$. As in the previous examples, this splits the path integral into a free Dirichlet-Dirichlet line, the detached vertex operator $V^{(r)}(T)\equiv-ig\,\phi^{(r)}(y)$ and the remaining semi-infinite line with parameter $\tau\in[T,\infty)$. Again, both the affine parameter and the insertion times $\tau_i$ of this Dirichlet-Neumann correlator can be shifted back to the standard $T$-independent range $[0,\infty)$. 

After cutting and gluing the path integral, the field $\phi^{(N+1)}$ is expressed as
\begin{equation}\label{phi n+1 right before cut and glue}
\begin{split}
\phi^{(N+1)}(x)&=-ig\,N!\int d^D y\int_0^\infty\!\!\! dT\, K_0(x,y;T)\Bigg\{ \frac{1}{N!}\,\phi^{(N)}(y)\prescript{}{y}{\!\Big\langle}\phi^{(1)}(X(\infty))\Big\rangle_{\rm DN}\\
    &+\sum_{r=1}^{N-1}\frac{1}{r!}\,\phi^{(r)}(y)\sum_{l=1}^{N-r}\frac{1}{l!}\sum_{m_1+\dots+m_l=N-r}\prescript{}{y}{\!\Big\langle}\prod_{i=1}^l  \frac{1}{m_i!}\overline{V}^{(m_i)}\phi^{(1)}(X(\infty))\Big\rangle_{\rm DN}\Bigg\}\;,    
\end{split}
\end{equation}
where in the sum over $r$ we separated the term $r=N$, which leaves no further insertions, and renamed $l$ as the number of remaining insertions, ranging now from $1$ to $N-r$ for fixed $r$. Now we can notice that this expression contains a lower order nonlinear field, since
\begin{align}
    \frac{1}{(N-r)!}\phi^{(N+1-r)}(y)=\sum_{l=1}^{N-r}\frac{1}{l!}\sum_{m_1+\dots+m_l=N-r}\prescript{}{y}{\!\Big\langle}\prod_{i=1}^l  \frac{1}{m_i!}\overline{V}^{(m_i)}\,\phi^{(1)}(X(\infty))\Big\rangle_{\rm DN}\;.
\end{align}
Substituting it in \eqref{phi n+1 right before cut and glue}, and recalling that $\prescript{}{y}{\!\Big\langle}\phi^{(1)}(X(\infty))\Big\rangle_{\rm DN}\equiv\phi^{(1)}(y)$, we obtain
\begin{align}
\label{substitute lower order}
    \phi^{(N+1)}(x)&=-ig\int d^D y\,Z_0(x,y) \sum_{r=1}^N \frac{N!}{r!(N-r)!}\,\phi^{(r)}(y)\phi^{(N+1-r)}(y)\;,
\end{align}
upon combining the term $r=N$ with the rest of the sum. Finally,  applying the wave operator we recover the equation of motion \eqref{phi eom binomial} for $\phi^{(N+1)}$:
\begin{align}
    \B \phi^{(N+1)}(x)&=g\sum_{r=1}^N\binom{N}{r}\phi^{(r)}(x)\phi^{(N+1-r)}(x)\equiv\frac{g}{2}\,\sum_{r=1}^N\binom{N+1}{r}\phi^{(r)}(x)\phi^{(N+1-r)}(x)\;,
\end{align}
where, to get to the second equality, we reshuffled the sum over $r$
\begin{equation}
\begin{split}
\sum_{r=1}^N\binom{N}{r}\phi^{(r)}\phi^{(N+1-r)}&=\sum_{s=1}^N\binom{N}{N-s+1}\phi^{(N-s+1)}\phi^{(s)}\equiv\sum_{r=1}^N\binom{N}{r-1}\phi^{(r)}\phi^{(N+1-r)}\;,    
\end{split}    
\end{equation}
and used Pascal's triangle identity
\begin{align}
 \binom{N}{r}+\binom{N}{r-1}=\binom{N+1}{r}\;.
\end{align}
With this we conclude the proof that, in $\phi^3$ theory, the worldline recursion \eqref{relation with exponential} builds nonlinear fields to all orders. Mapped to generalized vertex operators, they provide the worldline representation of tree-level amplitudes to all multiplicities.

\subsection{General scalar trees}
The picture that we introduced in the case of $\phi^3$ theory can be generalized to any polynomial scalar interaction. To illustrate how this generalization is achieved, we use the example of a massless theory with potential $\cU(\phi)=\frac{g}{3!}\,\phi^3+\frac{\lambda}{4!}\,\phi^4$. In this case, the equation of motion reads
\begin{equation}
    \label{phi 3 + phi 4 eom}
    \B\phi=\frac{g}{2}\,\phi^2+\frac{\lambda}{3!}\,\phi^3\;.
\end{equation}
To solve it, we make once again the ansatz \eqref{phi ansatz} for the field, which leads to the following recursive solution
\begin{align}
&\label{bilinearphi34}\phi^{(2)}=\frac{1}{\B}\,g\,\phi^{(1)}\phi^{(1)}\;, \\
&\label{multilinearphi34}\phi^{(n)} =\frac{1}{\B}\,\frac{g}{2}\sum_{k+l=n}\binom{n}{k}\phi^{(k)} \phi^{(l)} + \frac{1}{\B}\,\frac{\lambda}{3!}\sum_{k+l+m=n}\binom{n}{k,l,m}\phi^{(k)} \phi^{(l)} \phi^{(m)}\; ,\quad n\geq3\;,
\end{align}
in terms of the free field $\phi^{(1)}$. In \eqref{multilinearphi34} $\binom{n}{k,l,m}:=\frac{n!}{k!\,l!\,m!}$ is the coefficient of the trinomial expansion, meaning that $(a+b+c)^n\equiv \sum_{k+l+m=n}\binom{n}{k,l,m}a^kb^lc^m$. The nonlinear field $\phi^{(n)}$ is again given by the sum of unlabeled rooted trees with $n$ external legs. Compared to the pure cubic theory, the difference is that the diagrams contain both cubic and quartic vertices. The same ansatz \eqref{phi ansatz} can clearly be used for the case of arbitrary polynomial potentials.

From the worldline perspective, the crucial difference with the cubic theory is that vertex operators and fields are not simply proportional to each other as in \eqref{phi n and v n}. Let us illustrate how this situation now changes. At the linear level everything is the same, since the linear field $\phi^{(1)}(X(\tau))$ can only be mapped to $V^{(1)}(\tau)$ via \eqref{phi n and v n}. Similarly, the bilinear field $\phi^{(2)}$ can only be created with an asymptotic state at infinity and an insertion of a linear vertex operator. This is also evident from \eqref{bilinearphi34}, which only depends on the cubic coupling. For the nonlinear fields $\phi^{(n)}$ with $n\geq 3$, one has to start considering insertions of quartic vertices, as depicted in the following diagram for the three-particle field:
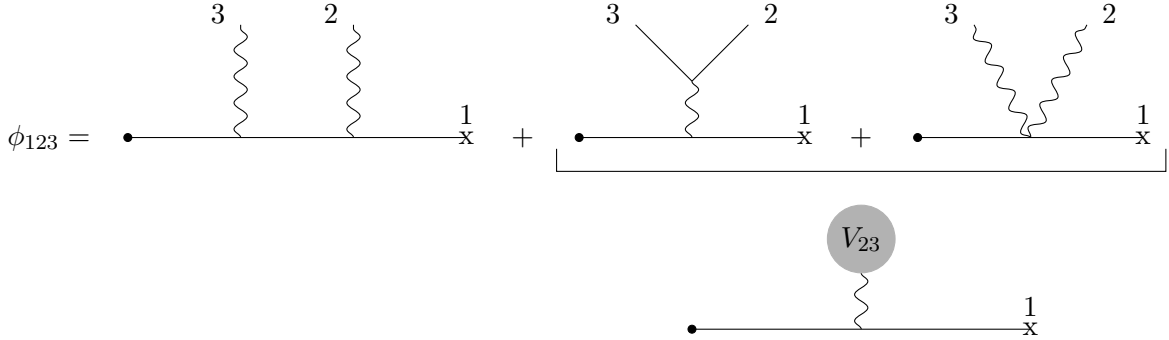
\begin{center}
    \begin{tikzpicture}[scale=1.5]
        \node at (0.3,0) {$\phi_{123}=$};
        \filldraw (1,0) circle (1pt);
        \draw (1,0)--(4,0);
        \draw[snake it] (2,0)--(2,1);
        \draw[snake it] (3,0)--(3,1);
        \node at (4,0) {x};
        \node at (4,0.2) {$1$};
        \node at (2.8,1.1) {$2$};
        \node at (1.8,1.1) {$3$};
        \node at (4.5,0) {$+$};

        \filldraw (5,0) circle (1pt);
        \draw (5,0)--(7,0);
        \draw[snake it] (6,0)--(6,0.5);
        \draw (6,0.5)--(5.5,1);
        \draw (6,0.5)--(6.5,1);
        \node at (7,0) {x};
\node at (7,0.2) {$1$};
\node at (6.7,1.1) {$ 2$};
\node at (5.3,1.1) {$ 3$};
        \node at (7.5,0) {$+$};
        
        \filldraw (8,0) circle (1pt);
        \draw (8,0)--(10,0);
        \draw[snake it] (9,0)--(8.5,1);
        \draw[snake it] (9,0)--(9.5,1);
        \node at (10,0) {x}; 
        \node at (9.7,1.1) {$ 2$};
        \node at (8.3,1.1) {$ 3$};
        \node at (10,0.2) {$ 1$};

        \draw(4.8,-0.1)--(4.8,-0.3)--(10.2,-0.3)--(10.2,-0.1);
        \filldraw (6,-1.7) circle (1pt);
        \draw (6,-1.7)--(9,-1.7);
        \draw[snake it] (7.5,-1.7)--(7.5,-1.2);
        \filldraw[gray!60] (7.5,-0.9) circle (0.3);
        \node at (7.5,-0.9) {$V_{23}$};
        \node at (9,-1.7) {x};
        \node at (9,-1.5) {$ 1$};
    \end{tikzpicture}   
\end{center}
As indicated in the picture, we can group the part of $\phi_{123}$ that comes from an insertion at a single point in the worldline, and use it to define a new bilinear vertex operator $V_{23}(\tau)$, related to the linear and bilinear fields by
\begin{equation}
\label{v12 phi3phi4}
    V_{12}(\tau)=-ig\,\phi_{12}(X(\tau))-i\lambda\,\phi_1(X(\tau))\,\phi_2(X(\tau))\;.
\end{equation}
For unpolarized vertex operators of arbitrary order, we infer that the map \eqref{phi n and v n} has to be generalized in the following way:
\begin{equation}
\label{phi 3+phi 4 vertex operators}
    V^{(n)}(\tau)=-ig\,\phi^{(n)}(X(\tau))-\frac{i\lambda}{2}\sum_{m+l=n}\binom{n}{m}\phi^{(m)}(X(\tau))\,\phi^{(l)}(X(\tau))\;,
\end{equation}
where $n\geq 2$. An additional symmetry factor of $\frac{1}{2}$ is now accompanying insertions of quadratic vertices, since they should be symmetric under the exchange of their indistinguishable inputs.

To obtain a multilinear field, one has to insert all possible lower order vertex operators into a worldline with a linear field state at infinity. These vertex operators can be computed by polarizing \eqref{phi 3+phi 4 vertex operators} according to the procedure we described in section \ref{sec:recursion}. To illustrate this, we depict graphically the computation of the four-particle field:

\begin{center}
    \begin{tikzpicture}[scale=1]
         \node at (0.7,0) {$\phi_{1234}(x)=$};
    \filldraw (2,0) circle (1pt);
    \draw (2,0)--(5,0);
    \draw[snake it] (2.5,1)--(2.5,0);
    \node at (2.2,1) {$4$};
    \draw[snake it] (3.5,1)--(3.5,0);
    \node at (3.2,1) {$3$};
    \draw[snake it] (4.5,1)--(4.5,0);
    \node at (4.2,1) {$2$};
    \node at (5,0) {x};
    \node at (5.2,0.3) {$1$};

    \node at (6,0) {$+$};

    \filldraw (7,0) circle (1pt);
    \draw (7,0)--(10,0);
    \draw[snake it] (8,0.5)--(8,0);
    \filldraw[gray!60] (8,1) circle (0.5);
    \node at (8,1) {$V_{43}$};
    \draw[snake it] (9,1)--(9,0);
    \node at (9.2,1.1) {$2$};
    \node at (10,0) {x};
    \node at (10.2,0.3) {$1$};
    
\node at (11,0) {$+$};
    \node at (12,0) {2 more};

\node at (1.5,-2) {$+$};
    \filldraw (2,-2) circle (1pt);
    \draw (2,-2)--(5,-2);
    \draw[snake it] (3.5,-2)--(3.5,-1.5);
    \draw (3.5, -1.5)--(3,-1);
    \draw (3.5,-1.5)--(4,-1);
    \draw (3.25,-1.25)--(3.5,-1);
    \node at (2.8,-0.7) {$4$};
    \node at (3.5,-0.7) {$3$};
    \node at (4.2, -0.7) {$2$};
    \node at (5,-2) {x};
    \node at (5.2,-1.7) {$1$};

    \node at (6,-2) {$+$};
    \node at (7,-2) {2 more};

    \node at (1.5,-4) {$+$};
    \filldraw (2,-4) circle (1pt);
    \draw (2,-4)--(5,-4);
    \draw[snake it] (3.5, -4)--(3.5,-3.5);
    \draw (2.5, -3)--(3.5,-3.5);
    \draw (3.5,-3.5)--(3.5,-3);
    \draw (3.5,-3.5)--(4.5,-3);
    \node at (2.5,-2.7) {$4$};
    \node at (3.5,-2.7) {$3$};
    \node at (4.5, -2.7) {$2$};
    \node at (5,-4) {x};
    \node at (5.2,-3.7) {$1$};

    \node at (6,-6) {$+$};
    \node at (7,-6) {2 more};

    \node at (1.5,-6) {$+$};
    \filldraw (2,-6) circle (1pt);
    \draw (2,-6)--(5,-6);
    \draw[snake it] (3.5, -6)--(3,-5.5);
    \draw (2.5, -5)--(3,-5.5);
    \draw[snake it] (3.5,-6)--(4.5,-5);
    \draw (3,-5.5)--(3.5,-5);
    \node at (2.5,-4.7) {$4$};
    \node at (3.5,-4.7) {$3$};
    \node at (4.5, -4.7) {$2$};
    \node at (5,-6) {x};
    \node at (5.2,-5.7) {$1$};

    \draw (8,-1.8)--(8.3,-1.8)--(8.3,-6.2)--(8,-6.2);

    \node at (8.8,-4) {$=$};
    \filldraw (9.2,-4) circle (1pt);
    \draw (9.2,-4)--(12.2,-4);
    \draw[snake it] (10.6,-4)--(10.6,-3);
    \filldraw[gray!60] (10.6,-3.2) circle (0.5);
    \node at (10.6,-3.2) {$V_{432}$};
    \node at (12.2,-4) {x};
    \node at (12.4, -3.7) {$1$};
    
    \end{tikzpicture}
\end{center}
We now point out that $\phi_{1234}$ is given by the same combination of generalized vertex operators as the one indicated in \eqref{phi1234}. We are thus led to propose that the relation \eqref{relation with exponential} holds even for a theory with both a cubic and a quartic coupling, provided that vertex operators are defined by the more general formula \eqref{phi 3+phi 4 vertex operators}.

We can once again prove that \eqref{relation with exponential} results in the same nonlinear fields as \eqref{multilinearphi34}. To this end, we repeat the cutting and gluing process that leads to \eqref{phi n+1 right before cut and glue}, from which we can obtain \eqref{substitute lower order} by noticing the presence of a lower order field given by \eqref{relation with exponential}. At that point, we have to use the new mapping \eqref{phi 3+phi 4 vertex operators} between vertex operators and fields and obtain the following relation: 
\begin{align}
\label{phi3 phi4 after cut and glue}
    \phi^{(N+1)}(x)=-\int d^Dy\,Z_0(x,y) &\left(ig\sum_{r=1}^N\binom{N}{r}\phi^{(r)}(y)\right.\notag
    \\&\left.+\frac{i\lambda}{2}\sum_{r=2}^{N}\sum_{p+q=r}\binom{N}{r}\binom{r}{p}\phi^{(p)}(y)\phi^{(q)}(y)\right)\phi^{(N+1-r)}(y)\;.
\end{align}  
Of the two terms, the first is identical to the case of $\phi^3$ theory and results in the bilinear term of the recursion \eqref{multilinearphi34}. In the second term, the sum over $r$ starts from $r=2$, because $p$, $q\geq 1$. 
The second sum in \eqref{phi3 phi4 after cut and glue} can be rewritten as follows:
\begin{align}
\label{rewrite trinomial 1}
    S_{3}&:=\sum_{r=2}^{N}\sum_{p+q=r}\binom{N}{N-r,p,q}\phi^{(p)}(y) \phi^{(q)}(y) \phi^{(N+1-r)}(y)\notag \\
    &=\sum_{p+q+r=N+1}\binom{N}{r-1,p,q}\phi^{(p)}(y) \phi^{(q)}(y) \phi^{(r)}(y)\;.
\end{align}
Thanks to the symmetry in the labels, we can split \eqref{rewrite trinomial 1} as
\begin{align}
    S_3=\frac{1}{3}\sum_{p+q+r=N+1}\left[\binom{N}{r-1,p,q}+\binom{N}{r,p-1,q}+\binom{N}{r,p,q-1}\right]\phi^{(p)}(y) \phi^{(q)}(y) \phi^{(r)}(y)\;,
\end{align}
which allows us to use Pascal's tetrahedron identity:
\begin{equation}
    \label{trinomial}
    \binom{N+1}{r,p,q}=\binom{N}{r-1,p,q}+\binom{N}{r,p-1,q}+\binom{N}{r,p,q-1}\;,
\end{equation}
valid for $r+p+q=N+1$. With this we finally recast \eqref{phi3 phi4 after cut and glue} in the form
\begin{align}
\label{phi3 phi4 final}
    \phi^{(N+1)}(x)&=-i\int d^Dy\, Z_0(x,y) \left(\frac{g}{2}\sum_{r+p=N+1}\binom{N+1}{r}\phi^{(r)}(y)\phi^{(p)}(y)\right.\notag \\&\left.+\frac{\lambda}{3!}\sum_{p+q+r=N+1}\binom{N+1}{r,p,q}\phi^{(p)}(y) \phi^{(q)}(y) \phi^{(r)}(y)\right),
\end{align} 
which coincides with \eqref{multilinearphi34}, as it can be seen by applying the wave operator.

We conclude by generalizing the above proof to arbitrary polynomial interactions. The method is identical: one starts from the cutting and gluing of the path integral formula \eqref{relation with exponential}, and makes use of recursion relations for multinomial coefficients, as given by Pascal's simplexes. More specifically, a monomial interaction $\frac{\lambda_n}{n!}\,\phi^n$ in the scalar potential corresponds to the contribution
\begin{equation}
\Delta V^{(r)}(\tau)=-\frac{i\lambda_n}{(n-2)!}\sum_{p_1+\dots+p_{n-2}=r}\binom{r}{p_1,\dots,p_{n-2}}\phi^{(p_1)}(X(\tau))\dots \phi^{(p_{n-2})}(X(\tau))\;,    
\end{equation}
to the generalized vertex operator $V^{(r)}(\tau)$, for $r\geq n-2$.
Upon reaching equation \eqref{phi3 phi4 after cut and glue}, the detached vertex operator leads to the additional term
\begin{align}
    S_{n-1}&:=\sum_{r=n-2}^N\sum_{p_1+\dots+p_{n-2}=r}\binom{N}{N-r,p_1,\dots,p_{n-2}}\phi^{(p_1)}(y)\dots \phi^{(p_{n-2})}(y) \phi^{(N+1-r)}(y)\notag \\
    &=\frac{1}{(n-1)}\sum_{p_1+\dots+p_{n-1}=N+1}\binom{N+1}{p_1,\dots,p_{n-1}}\phi^{(p_1)}(y)\dots \phi^{(p_{n-1})}(y)\;,
\end{align}
where we used Pascal's $(n-1)$-simplex recursion relation
\begin{align}
    \label{multinomial recursion}
    \binom{N+1}{p_1,\dots,p_{n-1}}=\sum_{m=1}^{n-1}\binom{N}{p_1,\dots,p_{m}-1,\dots,p_{n-1}}\;.
\end{align}
This term accounts for the part of $\phi^{(N+1)}$ originating from the $\phi^n$ vertex of the theory.
We have thus proven that the recursive relation \eqref{relation with exponential} holds for arbitrary polynomial interactions of scalar fields, the only difference being the relation between nonlinear fields and vertex operators.

\section{Conclusions and outlook}\label{sec:conclusions}

In this paper we have initiated a generalization of the standard techniques used in the worldline formalism, focusing on the case of self-interacting scalar theories. We have argued that the open line topology can be considered with three possible geometries, depending on the number of real (meaning finite distance) boundaries. Intriguingly, this classification strongly resembles the one adopted for intervals in the factorization algebra approach to quantum mechanics \cite{Costello:2021jvx,Chiaffrino:2024neo}. We have further demonstrated that asymptotic on-shell states can be consistently defined by inserting vertex operators at infinite times, and we have given a systematic derivation of the path integral for infinite and semi-infinite worldlines. We also introduced the notion of generalized vertex operators, representing the gluing of entire tree subdiagrams to a given worldline. Their correlation functions on infinite or semi-infinite lines generate tree-level amplitudes, or multiparticle fields, respectively. To make contact with standard QFT recursion relations for trees, we rewrote the worldline correlators in a way that makes a field theory interaction vertex manifest, by cutting open and re-gluing the path integral. Along these lines, it would be interesting to establish a precise prescription for gluing worldlines of different topologies, taking care of possible changes in the moduli space. This way, it might be possible to develop worldline master formulas for loop integrands.

The construction presented here is a first step towards an autonomous first-quantized representation of more general QFT processes, which until now required external inputs from the field theory itself. In particular, it would be beneficial to extend it to the case of Yang-Mills theory, whose first-quantized description has been given in \cite{Dai:2008bh,Bonezzi:2024emt,Bonezzi:2024fhd}. 
Since this more general framework allows one to exploit the full potential of the string-inspired formalism, it could provide valuable insights in making progress on the off-shell kinematic algebra of gauge theories \cite{Reiterer:2019dys,Ben-Shahar:2021zww,Borsten:2022vtg,Bonezzi:2022bse,Borsten:2023reb,Bonezzi:2023pox} and the related double copy program \cite{Bonezzi:2022yuh,Bonezzi:2022bse,Borsten:2023ned,Borsten:2023paw,Bonezzi:2024dlv,Bonezzi:2025anl}.

\section*{Acknowledgments}

We would like to thank Fiorenzo Bastianelli, Maor Ben-Shahar, Giuseppe Casale, Christoph Chiaffrino, Olindo Corradini, Filippo Fecit, Olaf Hohm and Jan Plefka for discussions and collaborations on closely related topics. The work of R.B. is funded by the Deutsche Forschungsgemeinschaft (DFG, German Research Foundation) – Projektnummer 524744955, “Worldline approach to the double copy”; M.F.K. is funded by the DFG – Projektnummer 417533893/GRK2575 “Rethinking Quantum Field Theory”.

\bibliography{Trees.bib}
\bibliographystyle{utphys}

\end{document}